\documentclass{aa}  

\usepackage{graphicx}
\usepackage{txfonts}
\usepackage[colorlinks=true, linkcolor=blue, citecolor=blue, filecolor=blue, urlcolor=blue]{hyperref}
\usepackage{booktabs}
\usepackage{multirow}
\usepackage{lscape}
\usepackage{threeparttable}

\newcommand{\teff}{$T_{\rm eff}$}
\newcommand{\logteff}{$\log(T_\mathrm{eff}/\mathrm{K})$}

\newcommand{\logg}{$\log(g)$}
\newcommand{\vmic}{$v_{\rm mic}$}

\newcommand{\vbroad}{$v_\mathrm{broad}$}
\newcommand{\logvbroad}{$\log(v_\mathrm{broad}/\mathrm{km\,s}^{-1})$}
\newcommand{\feh}{[Fe/H]}

\begin{document}

\title{A method to derive self-consistent NLTE astrophysical parameters for four million high-resolution 4MOST stellar spectra in half a day with invertible neural networks}
\titlerunning{Swift NLTE astrophysical parameters for 4 million high-res 4MOST spectra with INNs}

\author{Victor F.\ Ksoll
      \inst{1}
      \and
      Nicholas Storm\inst{2,3}
      \and
      Maria Bergemann\inst{2}
      \and
      Katherine Lee\inst{2}
      \and
      Ralf S.\ Klessen\inst{1,4}
      \and
      R. Albarracín\inst{2,3}
      \and 
      Guillaume Guiglion\inst{5, 2, 6}
      \and 
      Gra\v{z}ina Tautvai\v{s}ien\.{e}\inst{7}
      } 

\institute{Universit\"at Heidelberg, Zentrum f\"ur Astronomie, Institut f\"ur Theoretische Astrophysik, Albert-\"Uberle-Str. 2, 69120 Heidelberg, Germany\\
          \email{v.ksoll@uni-heidelberg.de}
     \and
     Max-Planck-Institut f\"ur Astronomie, K\"onigstuhl 17, 69117 Heidelberg, Germany
     \and
     Universit\"at Heidelberg, Grabengasse 1, 69117 Heidelberg, Germany
     \and 
     Universit\"{a}t Heidelberg, Interdisziplin\"{a}res Zentrum f\"{u}r Wissenschaftliches Rechnen, Im Neuenheimer Feld 225, 69120 Heidelberg, Germany
     \and 
     Zentrum f\"ur Astronomie der Universit\"at Heidelberg, Landessternwarte, K\"onigstuhl 12, 69117 Heidelberg, Germany 
     \and 
     Leibniz-Institut f{\"u}r Astrophysik Potsdam (AIP), An der Sternwarte 16, 14482 Potsdam, Germany
     \and
     Vilnius University, Faculty of Physics, Institute of Theoretical Physics and Astronomy, Sauletekio av. 3, 10257 Vilnius, Lithuania
         }

\date{Received 16 December 2025; accepted 19 February 2026}

  \abstract
   {Modern spectroscopic surveys have the capacity to obtain the spectra of millions of stars. However, classical spectroscopic methods can often be computationally expensive, rendering them impractical for the analysis of large datasets.}
   {We introduce a novel simulation-based, deep-learning approach for the efficient analysis of high-resolution stellar spectra that will be obtained with the upcoming high-resolution 4MOST spectrograph.}
   {We used a suite of synthetic non-local thermodynamic equilibrium (NLTE) spectra generated with Turbospectrum to mimic 4MOST observations and trained a conditional invertible neural network (cINN) for the purpose of predicting self-consistently stellar surface parameters and chemical abundances. The cINN is a neural network architecture that estimates full posterior distributions for the target stellar properties, providing an intrinsic uncertainty estimate.
   We evaluated the predictive performance of the trained cINN model on both synthetic data and the observed spectra of stars.}
  {We found that our new cINN trained on NLTE synthetic spectra is capable of recovering stellar parameters with average errors ($\sigma$) of $33$ K for \teff, $0.16$ dex for \logg, and $0.12$ dex for [Fe/H], $0.1$ dex for [Ca/Fe], $0.11$ for [Mg/Fe], and $0.51$ dex for [Li/Fe], respectively, at a signal-to-noise ratio (S/N) of 250 per Angstrom. From the analysis of the observed spectra of Gaia-ESO / 4MOST / PLATO benchmark stars, we verified that our NLTE estimates for stellar parameters and abundances are consistent with results obtained with the independent code TSFitPy. We conclude that the NLTE cINN is robust and that it can, in theory, evaluate four million high-resolution 4MOST spectra in less than a day, using GPU acceleration.}
  {}
   \maketitle
\nolinenumbers
%
\section{Introduction}
\label{sec:intro}
The past decade has seen revolutionary advances in observational stellar and Galactic astronomy. Large-scale stellar spectroscopic surveys, such as Gaia \citep{Gaia2016} and Gaia-ESO \citep{Gilmore2012GaiaESO, Gilmore2022, Randich2013GaiaESO, Randich2022}, GALAH \citep{DeSilva2015GALAH}, LAMOST \citep{Zhao2012LAMOST}, and APOGEE \citep{Majewski2016APOGEE, Majewski2017APOGEE} have provided  millions of stellar spectra, and upcoming facilities such as WEAVE \citep{Jin2024WEAVE}, 4MOST \citep[4-metre Multi-Object Spectroscopic Telescope;][]{DeJong2019_4MOST}, and SDSS-V \citep{Kollmeier2026SDSS_V} will further increase the data quantity by several orders of magnitude. 

To make use of this enormous wealth in observational data, all of these spectroscopic datasets require efficient analysis approaches. Classic fitting tools such as MOOG \citep{Sneden1973}, SIU \citep{Reetz1991}, PySME \citep{Wehrhahn2023}, KORG \citep{Wheeler2023}, and TSFitPy \citep{Gerber2023, Storm2023} are highly flexible in terms of micro-physics. However, these codes typically require hours of CPU time in order to provide a detailed analysis of chemical abundances and stellar parameters for a single star. Although this approach is highly desirable and justified for individual stars, it becomes prohibitively expensive in terms of computation time when homogeneous, precise, and fast analyses of millions of high-resolution stellar spectra are required.

Consequently, approaches that employ machine learning have been gaining considerable traction in recent years as a more efficient alternative. In particular, neural network-based methods have proven to be promising. \cite{Ting2019} introduced a spectral fitting tool known as 'the Payne', which uses a simple feed-forward neural network to facilitate a precise interpolation of synthetic spectra in a very high-dimensional parameter space. It was applied, for instance, in the analysis of GALAH DR4 \citep{Buder2025}.
Another data-driven spectral analysis approach is known as 'the Cannon' \citep{Ness2015}, which uses real, well-analysed spectra to build an efficient generative model for the probability distribution of the flux per wavelength given the stellar parameters. It was adapted for APOGEE spectra in \citet{Ness2016a}. The application of convolutional neural networks (CNNs) has also been explored in, for instance, the analysis of RAVE spectra \citep{Guiglion2020} and Gaia RVS spectra \citep{Guiglion2024}. 
The analysis of spectra has also seen exploration of various generative deep learning approaches. \cite{Liu2024_StellarGAN} introduced the StellarGAN, a spectral classification approach for O to M-type stars based on generative adversarial networks \citep[GANs;][]{GoodfellowNIPS2014} that is trained on SDSS and APOGEE data. Another GAN-based approach is the GANDALF framework \citep{Santovena2024, Manteiga2025}, which is trained on APOGEE and GALAH spectra and employs an autoencoder to embed spectra in a lower dimensional representation space prior to deriving stellar atmospheric properties. More recently, \cite{Cvrcek2025} presented a method that uses variational autoencoders \citep[VAEs;][]{Kingma2013, Rezende2014} to both predict stellar parameters and simulate spectra, which they trained on both real and synthetic spectra from the HARPS \citep{Mayor2003} instrument. \cite{Gebran2025} also employed conditional VAEs \citep[cVAEs;][]{Sohn2015} to build a physics-aware surrogate model for spectral synthesis, trained on synthetic spectra from SYNSPEC \citep{Hubeny2011}. However, most of these approaches rely on stellar parameters and chemical abundances, obtained using highly simplified stellar synthetic spectra computed under assumptions of local thermodynamic equilibrium (LTE). As demonstrated in numerous recent studies \citep[e.g.][]{Korn2003, Asplund2005, Caffau2010, Lind2011, Bergemann2012a, Bergemann2012b, Asplund2021, Amarsi2020, Lind2024, Bergemann2025}, derivation of precise and accurate stellar parameters and abundance requires that stellar spectroscopic model grids take into account the effects of non-LTE (NLTE).

Here, we present a novel approach to provide accurate astrophysical characterisation for gigantic sets of data; specifically, from the upcoming 4MOST high-resolution disc and bulge survey 4MIDABLE-HR \citep{Bensby2019}. In this survey (PI: Bensby T. and Bergemann M.),  around four million spectra will be obtained at high resolution ($\mathrm{R}=20\,000$) to study the evolution of the Galactic disc and bulge. 
Dictated by the need to have highly efficient, stable, and adaptable algorithms, as well as the need to include NLTE in the chemical analysis \citep{Lind2024, Bergemann2025}, our aim is to expand upon the code presented in \citet{Kovalev2019} and applied to the Gaia-ESO survey therein. 
In particular, we introduced a novel simulation-based inference approach that for the first time combines NLTE synthetic spectra with a conditional invertible neural network \citep[cINN;][]{Ardizzone2019a, Ardizzone2019b}. Overall, 
cINNs are a deep learning architecture that belongs to the generative normalising flow approaches and are specifically tailored towards solving inverse problems. They estimate the full posterior distribution, providing both an intrinsic uncertainty measure for the predictions and highlighting potential degeneracies in the given inverse problem. 
Although VAEs and GANs can be employed in a similar fashion \citep[as in e.g.~][]{Cvrcek2025}, INNs have been shown to return better calibrated posterior distributions \citep{Ardizzone2019a} and to be more robust to mode collapse \citep{Ardizzone2019b, Kobyzev2021, BondTaylor2022}.
In addition, cINNs have already seen a series of successful applications in astronomy. These include the characterisation of stars from photometry \citep{Ksoll2020},  analysis of emission lines from HII regions \citep{Kang2022, Kang2023a},  inference of cosmic ray origins \citep{Bister2022},  evaluation of exoplanet observations \citep{Haldemann2023},  prediction of galaxy merger histories \citep{Eisert2023}, and  3D reconstruction of dust distributions in star-forming cores \citep{Ksoll2024}.
In the role that is most relevant to the work presented in this study, cINNs have been proven to work well for the analysis of stellar spectra in several applications. This includes the evaluation of MUSE spectra based on LTE stellar atmosphere models from the PHOENIX library \citep{Kang2023b, Kang2025} and the characterisation of high resolution spectra using a cINN trained on observed spectra of Milky Way field and cluster stars obtained with the GIRAFFE spectrograph in the Gaia-ESO survey \citep{Candebat2024}.

In the following, we outline our development of the first self-consistent cINN method for the analysis of observations of stars with the high-resolution spectrograph of 4MOST that is based fully on 25-D NLTE stellar spectral model grids. 
In Sect.~\ref{sec:data}, we introduce the training data for our cINN model, as well as a set of real benchmark spectra that we use to ascertain the performance of our model. Section~\ref{sec:method} outlines the basic concepts of the cINN approach and implementation details for the specific inverse problem that we aim to solve in this study. In Sect.~\ref{sec:results}, we then describe the performance of our approach on both synthetic test data, as well as on archival spectra of the Gaia-ESO/4MOST/PLATO benchmark stars. Lastly, Sect.~\ref{sec:summary} provides a summary of our main results and outlook for the subsequent development of our method.

\section{Data}
\label{sec:data}

In this section, we briefly outline the preparation of the training data (Sect.~\ref{sec:training_data}). We also describe the composition of the set of real benchmark spectra used in Sect.~\ref{sec:results_benchmark} to gauge the performance of our trained model on real data (Sect.~\ref{sec:data_benchmark}).

\subsection{Training data}
\label{sec:training_data}

\begin{figure*}
    \centering
    \includegraphics[width=0.7\linewidth]{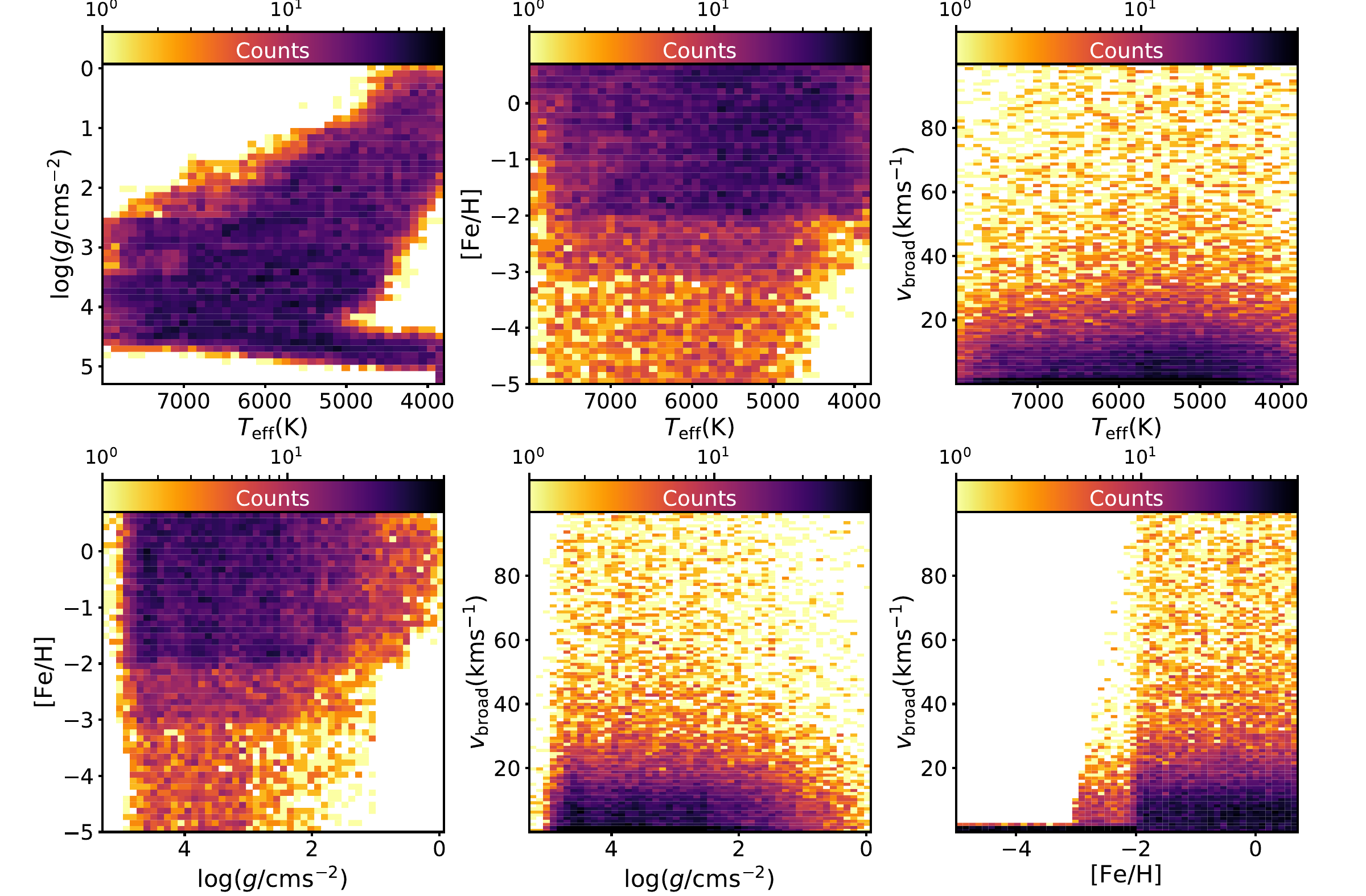}
    \includegraphics[width=0.99\linewidth]{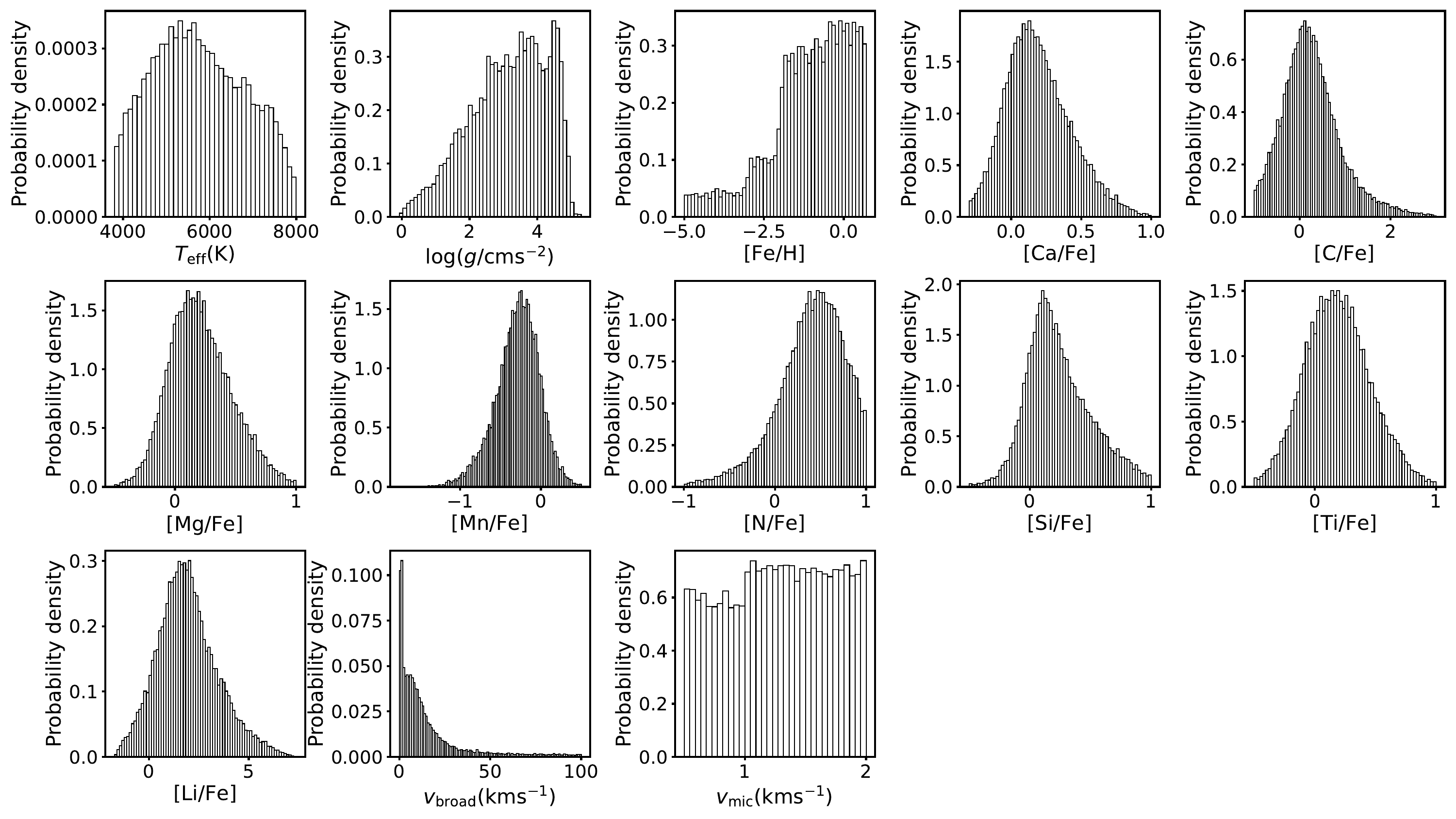}
    \caption{Prior distributions of the target parameters in the training data. The panels in the top two rows show 2D histograms of the correlation between selected stellar properties. The panels in the bottom three rows show the 1D histograms of all target parameters.}
    \label{fig:ts_priors}
\end{figure*}

We computed large grids of NLTE spectra (29\,548 stellar spectral models) with our latest version of Turbospectrum NLTE \citep[TSv20\footnote{\url{https://github.com/bertrandplez/Turbospectrum_NLTE}};][]{Gerber2023, Storm2023}, using the TSFitPy Python wrapper\footnote{\url{https://github.com/TSFitPy-developers/TSFitPy}}.
The generated spectra employed 1D MARCS model atmospheres \citep{Gustafsson2008} and cover a wavelength range of 3670 to 9800~\AA~in steps of 0.007~\AA~with perfect resolution without any broadening, as well as varying stellar parameters (\teff, \logg, [Fe/H]) and abundances for 21 elements (Li, C, N, O, Na, Mg, Al, Si, Ca, Ti, V, Cr, Mn, Co, Ni, Sr, Y, Zr, Ba, Ce, Eu). 
In our spectrum synthesis calculations, we adopted a random distribution of micro-turbulence parameter values  \vmic, ranging from 0.5 to 2 km\,s$^{-1}$. We used the Gaia-ESO line lists \citep{Heiter2021} with new atomic data for C, N, O, Si, Mg, as described in \citep{Magg2022}, and VALD for its gaps \citep{Ryabchikova2015}. The chemical composition of the Sun is adopted from our NLTE analysis in \cite{Magg2022}. 
There were 16 elements simulated in NLTE: H \citep{Mashonkina2008}, O \citep{Bergemann2021}, Na \citep{Ezzeddine2018}, Mg \citep{Bergemann2017}, Al \citep{Ezzeddine2018}, Si \citep{Bergemann2013, Magg2022}, Ca \citep{Mashonkina2017, Semenova2020}, Ti \citep{Bergemann2011}, Mn \citep{Bergemann2019}, Fe \citep{Bergemann2012a, Semenova2020}, Co \citep{Bergemann2010b, Yakovleva2020}, Ni \citep{Bergemann2021, Voronov2022}, Sr \citep{Bergemann2012b, Gerber2023}, Y \citep{Storm2023, Storm2024}, Ba \citep{Gallagher2020}, and Eu \citep{Storm2024}. Subsequently, we convolved the synthetic stellar spectra with a Gaussian kernel to get a resolving power of $R= 80\,000$. We interpolated the spectra onto a grid with a pixel size of $0.025\,\text{\AA}$. 
Afterwards, rotational broadening (\vbroad) was applied, sampled between 0 and 100 km\,s$^{-1}$. We note that we opted to not perform the immediate convolution with the 4MOST HR resolution so that we  could retain the capability to test these grids on spectra with higher resolving power (also those analysed as part of 4MOST work, e.g. the Gaia-ESO benchmark stars, which are available at R$\gtrsim 47\,000$). Therefore only in the final step, we '4MOST-ified' the spectra by degrading them to R of the 4MOST high-resolution spectrograph ($\mathrm{R} = 20\,000$) and to different signal-to-noise  ratios (S/Ns) using the 4FS ETC package\footnote{\url{https://escience.aip.de/readthedocs/OpSys/etc/master/index.html}}. 
The synthetic spectra cover the full range of FGK-type stars in the \teff-log(g) diagram with [Fe/H] from $-$5 to $+$0.5. In total, we generated a base set of $29\,548$ model NLTE spectra. 
Figure~\ref{fig:ts_priors} shows the prior distributions of the relevant stellar surface parameters (i.e.~\teff, \logg, \feh, \vbroad, \vmic, [Ca/Fe], [C/Fe], [Mg/Fe], [Mn/Fe], [N/Fe], [Si/Fe], [Ti/Fe], and [Li/Fe]) considered in the training of our method across this base set. We note that the break in \vbroad~at around $\mathrm{[Fe/H]} = -2$ in this model grid is based upon the assumption that old stars rotate more slowly overall. 
We then randomly split this database into 75\%, 20\% and 5\% sub-samples, dedicated to training, test, and validation, respectively.
For each of these spectra, we then computed versions with 19 different S/N values (namely, 10, 12, 14, 16, 18, 20, 23, 26, 30, 35, 40, 45, 50, 80, 100, 130, 180, 250, and 1000 \AA$^{-1}$), motivated by the need to cover the entire dynamic range of spectral data quality potentially available with 4MOST (e.g. bright calibration stars versus faint distant targets in the deep bulge or disc-halo interface). 
After accounting for this data augmentation step, our final model spectra database across training, test, and validation sets consisted of $561\,412$ spectra. 

\subsection{Benchmark data}
\label{sec:data_benchmark}

As 4MOST is not yet operational, we cannot benchmark our approach on real 4MOST data. To get an idea of how the cINN performs on observed spectra, we therefore used archival spectra of selected well-studied benchmark stars in addition to some other supplementary bright targets from our previous works. The spectra of the Gaia-ESO  benchmark stars \citep{Jofre2014, Heiter2015} were taken from  \citet{BlancoCuaresma2014}. These stars were observed with different astronomical facilities, including the Ultraviolet and Visual Echelle Spectrograph (UVES) at the 8m Very Large Telescope (VLT) of the European Southern Observatory (ESO) in Chile and the Narval spectrograph at the 2.03m Telescope Bernard Lyot (TBL) on the Pic du Midi de Bigorre in France. We note that some of the UVES spectra stem from the UVES Paranal Observatory Project \citep[UVES POP,][]{Bagnulo2003}. Other stars were analysed in \citet{Fuhrmann1993, Fuhrmann1998, Gehren2004, Gehren2006, Bergemann2008, Bergemann2010a, Bergemann2011}, based on the spectroscopic data collected with the fibre optics Cassegrain \'echelle spectrograph (FOCES) at the 2.2m Calar-Alto telescope.
Table~\ref{tab:benchmark_params} in the appendix lists the set of sources that we selected together with literature results for \teff, \logg~and \feh~(with the corresponding references). In total, our test set consisted of 58 unique stars across 72 observed spectra, including 39 main sequences stars, 11 subgiants, and 8 giants from the K to F types. All the spectra were continuum normalised and corrected for radial velocity. 

All of the benchmark spectra have a higher native resolution than 4MOST, existing on a different wavelength grid than the output of the 4MOST high resolution spectrograph and have varying S/Ns. To process them with the cINN approach built in this work, we had to first 4MOST-ify the spectra. This included a reduction of the resolution to $\mathrm{R} = 20\,000$ (matching the target resolution of the 4MIDABLE-HR survey) via convolution with an appropriate Gaussian kernel and rebinning the spectra to the wavelength bin grid of the high resolution 4MOST spectrograph. As most of the spectra in our test set do not extend far enough into the blue to cover the blue arm ($3926$ to $4355$ \AA) of the 4MOST high resolution spectrograph, the latter rebinning step was limited to the green ($5160$ to $5730$ \AA) and red  ($6100$ to $6790$ \AA) arms only. Consequently, the network presented in this study, operates only on these two wavelength windows. 
While the cINN described further below was trained to handle different S/Ns in the input data, it is limited to the discrete set of S/N values currently generated in the training data (see Sect.~\ref{sec:training_data}). For the purposes of the benchmark conducted in Sect.~\ref{sec:results_benchmark}, we decided to degrade all spectra to a fixed $(\mathrm{S/N})_\mathrm{fixed} = 250\,$\AA$^{-1}$. All spectra considered have a native S/N that is higher, so that degrading the S/N (per wavelength bin) to that value was straightforward via the addition of Gaussian noise following
\begin{equation}
    f_\mathrm{i, pert} = f_\mathrm{i} + \Delta f_\mathrm{i}, \, \mathrm{with}\, \Delta f_\mathrm{i} \propto \mathcal{N}\left(\mu = 0, \sigma = f_\mathrm{i} \times (\mathrm{S/N})_\mathrm{fixed}^{-1}\right),
\end{equation}
where $f_\mathrm{i}$ denotes the flux in wavelength bin $i$ at the native S/N of the spectrum. 

\section{Method}
\label{sec:method}

The following provides a short overview of the cINN methodology (Sect.~\ref{sec:cinn_intro}), the specific architecture and implementation choices for the model used in this work (Sect.~\ref{sec:architecture_details}), and the network training strategy (Sect.~\ref{sec:training_strategy}). In addition, we detail our procedure for hyperparameter optimisation (Sect.~\ref{sec:hyperparameters}) and the post-processing procedure we applied to derive point estimates from the predicted posterior distributions (Sect.~\ref{sec:point_estimation}). 

\subsection{The conditional invertible neural network}
\label{sec:cinn_intro}
Normalising flows \citep[NFs,][]{Tabak2010, Tabak2013, Dinh2015, Rezende2015, Kobyzev2021} are a family of generative deep learning models that excel at modelling complex distributions. To do so, NFs map simpler known probability distributions (that can be easily sampled from) to the complex target distribution by employing a series of invertible transformations \citep[see e.g.][for a review]{Kobyzev2021}. In the following, we utilise a variant of the NF framework, called the conditional invertible neural network \citep[cINN,][]{Ardizzone2019b}. Among the NFs, the cINN architecture is particularly well suited for solving inverse problems. The latter comprise tasks, where it is the goal to infer the fundamental (physical) properties $\mathbf{x}$ of some system (e.g.~a star) given a corresponding set of observable quantities, $\mathbf{y}$. Inverse problems are ubiquitous in the natural sciences, but solving them (i.e. recovering the mapping $\mathbf{y} \rightarrow \mathbf{x}$) is often challenging. One of the main difficulties arises from the fact that for many problems in nature the forward process $\mathbf{x} \rightarrow \mathbf{y}$ that links the (unobservable) properties of the system to the actual observables suffers from an inherent loss of information. As a consequence, different combinations of the system properties can result in similar or even identical observations, rendering the inverse problem degenerate. 

To tackle these challenges, the cINN uses a latent space approach to capture the information lost in the mapping from physical properties to actual observables. More specifically, instead of directly learning the inverse mapping $\mathbf{y} \rightarrow \mathbf{x}$, the cINN learns a mapping, $f$, of the physical properties, $\mathbf{x}$, to the latent variables, $\mathbf{z}$, conditioned on the observables, $\mathbf{y}$, that is, 
\begin{equation}
    \mathbf{z} = f(\mathbf{x}; \mathbf{c} = \mathbf{y}).
\end{equation}
This training strategy employs the latent space to encode variance of the target parameters, $\mathbf{x}$, that is not captured by the corresponding observables, while ensuring that a prescribed prior distribution $P(\mathbf{z})$ for the latent variables is maintained during training. In this context, $P(\mathbf{z})$ is usually selected to be a multivariate normal distribution with zero mean and unit covariance and the dimension of the latent space matches that of the target parameter space per construction of the mapping. After a cINN is trained in this manner, given a new observation $\mathbf{y}'$ the cINN can then query the variance encoded in the latent space via sampling from the known prior $P(\mathbf{z})$ and, leveraging the invertible architecture of the cINN, compute samples of the posterior distribution $p(\mathbf{x}|\mathbf{y}')$ following
\begin{equation}
    p(\mathbf{x}|\mathbf{y}') \sim g(\mathbf{z}; \mathbf{c} = \mathbf{y}') \,\,\,\mathrm{with}\,\,\, \mathbf{z} \propto P(\mathbf{z}),
\end{equation}
where $g(\cdot;\mathbf{c}) = f^{-1}(\cdot;\mathbf{c})$ denotes the inverse of the forward mapping, $f$, for fixed condition, $\mathbf{c}$.

The fundamental architecture of the cINN as described above directly constrains the dimension of the latent space. However, it does not impose any such limitation on the dimension of the input feature space as the observations enter only as a conditioning input in the learned mappings $f$ and $g$ of the cINN. This not only allows $\dim(\mathbf{y})$ to be arbitrarily large, but also permits the addition of a secondary feature extraction network in a cINN. This feature extraction network, usually trained in tandem with the main cINN, can then transform the raw input observations into a more useful (learned) representation (usually lower dimensional), which can improve the predictive performance of a cINN \citep{Ardizzone2019b}. The cINN itself is comprised of a series of invertible, conditional affine coupling blocks \citep{Dinh2017, Ardizzone2019b}. Given an input vector, $\mathbf{u}$, each of these coupling blocks computes two complementary affine transformations with element-wise multiplication, $\odot$, and addition, $\oplus$, namely
\begin{equation}
    \label{eq:coupling_forward}
    \begin{split}
        \mathbf{v}_1 &= \mathbf{u}_1 \odot \exp\left(s_2(\mathbf{u}_2, \mathbf{c})\right) \oplus t_2(\mathbf{u}_2, \mathbf{c}), \\
        \mathbf{v}_2 &= \mathbf{u}_2 \odot \exp\left(s_1(\mathbf{v}_1, \mathbf{c})\right) \oplus t_1(\mathbf{v}_1, \mathbf{c}), 
    \end{split}
\end{equation}
where $\mathbf{u}_1$, $\mathbf{u}_2$ and $\mathbf{v}_1$, $\mathbf{v}_2$ denote halves of the input $\mathbf{u}$ and output vector $\mathbf{v}$, respectively. In these transformations, $s_i$ and $t_i$ represent arbitrarily complex mappings of the concatenation of $\mathbf{u}_i$/$\mathbf{v}_i$ and the conditioning input, $\mathbf{c}$ (i.e.~the input observations or output of a feature extraction network if present). From Eq.~\eqref{eq:coupling_forward} we can see that the full transformation carried out by the coupling block is easily inverted given the output vector, $\mathbf{v} = (\mathbf{v_1}, \mathbf{v_2})$, following 
\begin{equation}
    \label{eq:coupling_backward}
    \begin{split}
        \mathbf{u}_2 &= \left(\mathbf{v}_2 \ominus t_1(\mathbf{v}_1, \mathbf{c})\right) \odot \exp\left(-s_1(\mathbf{v}_1, \mathbf{c})\right), \\
        \mathbf{u}_1 &= \left(\mathbf{v}_1 \ominus t_2(\mathbf{u}_2, \mathbf{c})\right) \odot \exp\left(-s_2(\mathbf{u}_2, \mathbf{c})\right),
    \end{split}
\end{equation}
where $\ominus$ denotes element-wise subtraction. The mappings $s_i$ and $t_i$ introduced in Eqs.~\eqref{eq:coupling_forward} and \eqref{eq:coupling_backward} are never inverted themselves, regardless of whether the forward or backward passes of the coupling block are computed. This means that $s_i$ and $t_i$ are not required to be invertible transformations themselves. In fact, these mappings do not even need to be prescribed before the cINN is trained, but can be represented by small sub-networks instead (e.g.~a simple fully connected feed forward network), such that $s_i$ and $t_i$ are learnt during training of the cINN \citep{Ardizzone2019a, Ardizzone2019b}.

\subsection{Architecture and implementation details}

\begin{figure*}
    \centering
    \includegraphics[width=0.9\textwidth]{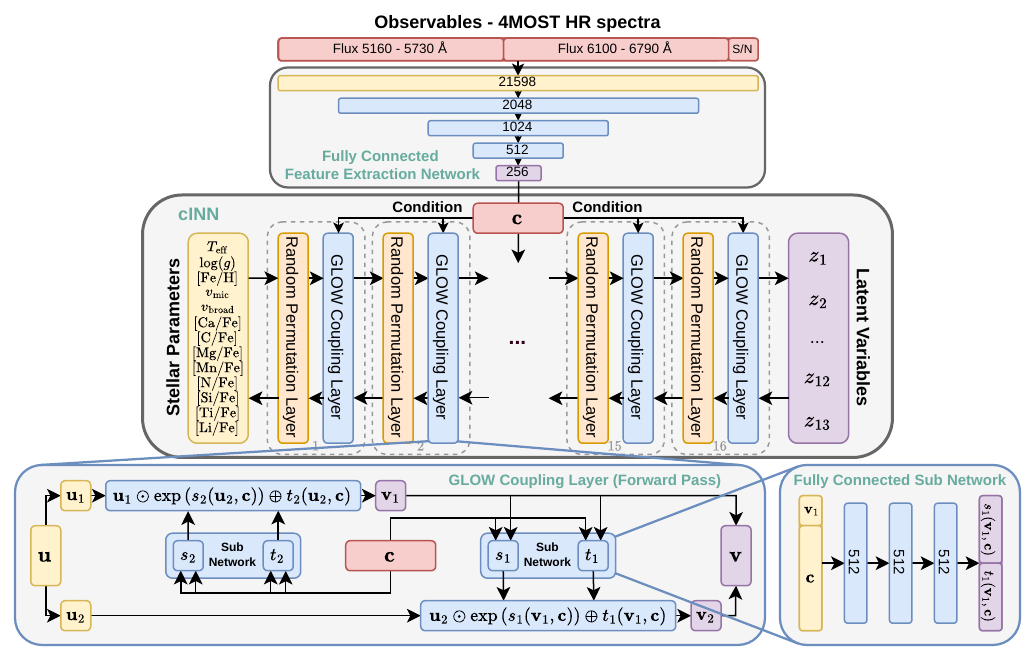}
    \caption{Schematic overview of the cINN architecture. We note that the bottom zoom-in only highlights the forward pass of a GLOW coupling layer, following Eq.~\eqref{eq:coupling_forward}, but not the backward pass (Eq.~\eqref{eq:coupling_backward}). Each of the 16 GLOW coupling blocks in our architecture employs two sub-networks, following the layout shown in the bottom right zoom-in.}
    \label{fig:cINN_architecture}
\end{figure*}

\label{sec:architecture_details}
Because most of the observed spectra in our set of benchmark observations (see Sect.~\ref{sec:data_benchmark}) do not extend far enough to cover the blue arm ($3926$ to $4355$ \AA) of the high-resolution 4MOST spectrograph, the model we trained in the following used only the fluxes from the green ($5160$ to $5730$ \AA) and red arm ($6100$ to $6790$ \AA). 
An extension of our method to the blue arm once actual 4MOST spectra are available is certainly straightforward and preliminary tests on synthetic spectra indicate that the usage of the full 4MOST HR coverage will result in an even better performance of our approach. 
In addition to the fluxes, we also adopted the average S/N as a model input, so that the trained cINN can account for the uncertainty in the input fluxes in the predicted posteriors, as previously demonstrated in \cite{Kang2023a, Kang2023b, Kang2025}. 
For simplicity, we further made sure that the input spectra to the cINN are both continuum normalised, as well as corrected for radial velocity. After some initial experimentation, we identified the following target parameters as viable to recover with the cINN given the input wavelength coverage (as per tests on synthetic data): \teff, \logg, \feh, \vmic, \vbroad, [Ca/Fe], [C/Fe], [Mg/Fe], [Mn/Fe], [N/Fe], [Si/Fe], [Ti/Fe], and [Li/Fe]. As both \teff~and \vbroad~cover a rather large range, the network is trained to predict \logteff~and \logvbroad instead. This procedure may avoid training and convergence issues that can occur when the target parameters span vastly different orders of magnitude. In addition, both the input fluxes (per wavelength bin) and all target parameters are also standardised across the training data. Standardisation refers here to the practice of subtraction of the mean and subsequent normalisation by the inverse standard deviation.  

To implement the cINN for the purposes of this study, we employed the dedicated Framework for Easily Invertible Architectures \citep[FrEIA\footnote{Available at \url{https://github.com/vislearn/FrEIA}},][]{Ardizzone2019a, Ardizzone2019b} package, based on the {\sc pytorch} \citep{Paszke2017_pytorch} Python deep learning module. 
Furthermore, we adopted the Generative Flow \citep[GLOW;][]{Kingma2018} architecture for the affine coupling blocks, which introduces only two sub-networks per coupling block, jointly modelling $(s_1, t_1)$ and $(s_2, t_2)$ by one sub-network each, to improve computational efficiency. For the set-up of the sub-networks themselves, we selected a simple, fully connected network design that consists of three layers of size $512$ with rectified linear unit (ReLU) activation functions. 
\citet{Ardizzone2019b} note that the exponential term $\exp(s)$ in Eqs.~\eqref{eq:coupling_forward} and \eqref{eq:coupling_backward} could lead to instabilities during training of a cINN due to an exploding magnitude problem. To mitigate this issue, we also adapted the clamping procedure proposed in \citet{Ardizzone2019b}, which modifies the argument, $s$, of the exponential function according to
\begin{equation}
    s_{\mathrm{clamp}} = \frac{2\alpha}{\pi} \arctan\left(\frac{s}{\alpha}\right),
\end{equation}
where $\alpha = 1.9$.
In addition, we introduced random permutation layers before each affine coupling block, which permute the input vector in a random but fixed (and therefore invertible) manner. The inclusion of these permutation layers serves as a way to intermix the information between the two streams $\mathrm{u}_1$ and $\mathrm{u}_2$ of the coupling blocks.
As the input feature space is rather high dimensional given the high resolution of the 4MOST spectra, we also decided to add a feature extraction network to our cINN, so that the cINN can extract worthwhile features from the input data before passing them to the main body of the network. Specifically, we used a fully connected network with three hidden layers of size 2048, 1024, and 512, each employing the ReLU activation function, which returns 256 features in the output layer. 
We note that in preliminary experiments we also tested small convolutional neural networks for feature extraction in a cINN, but found that they did not yield significant performance improvements over the simpler fully connected architecture, while at the same time slowing down the inference. 
As described in Sect.~\ref{sec:cinn_intro}, the feature extraction network is trained together with the main cINN itself on the same loss function. In our final network set-up (as determined by hyperparameter optimisation, see Sect.~\ref{sec:hyperparameters}), the cINN itself consisted of $16$ affine GLOW coupling blocks. A schematic overview of the full architecture is given in Fig.~\ref{fig:cINN_architecture}.

\subsection{Training set-up}
\label{sec:training_strategy}
To train the cINN, we followed the approach in \cite{Ardizzone2019b} and employed the maximum likelihood loss, 
\begin{equation}
    \mathcal{L} = \mathbb{E}_i \left(\frac{||f(\mathbf{x}_i; \mathbf{c}_i, \Theta)||_2^2}{2} - J_i \right),
\end{equation}
where $J_i$ is the determinant of the Jacobian matrix $J_i = \det(\partial f/\partial \mathbf{x}|_{\mathbf{x} = \mathbf{x}_i})$, evaluated at training example $\mathbf{x}_i$, as the primary optimisation goal for the network weights $\Theta$. In addition, we imposed a secondary loss $\mathcal{L}_\mathrm{rev}$ for the backward pass of the cINN. Specifically, after the training example $(\mathbf{x}_i, \mathbf{c}_i)$ has passed through the cINN in the forward direction, that is $\mathbf{z_i} =  f(\mathbf{x}_i; \mathbf{c}_i, \Theta)$, in the computation of $\mathcal{L}$, we can derive the secondary loss, which is also to be minimised as
\begin{equation}
    \mathcal{L}_\mathrm{rev} = \mathbb{E}_i \left[\mathbf{x}_i - g(\mathbf{z}_i + \delta \mathbf{z}_i; \mathbf{c}_i, \Theta)\right]^2,
\end{equation}
where $\delta\mathbf{z}_i$ denotes a small amount of Gaussian noise, that is $\delta \mathbf{z}_i \sim \mathcal{N}(\mathbf{\mu} = \mathbf{0}, \Sigma = 10^{-5} \times \mathbb{I})$. Here, $\mathcal{L}_\mathrm{rev}$ is supposed to promote a more robust embedding in the latent space, such that the invertibility of the network holds even when subject to small perturbations in the latent space. In practise, $\mathcal{L}_\mathrm{rev}$ drops significantly faster than $\mathcal{L}$, however, so that the maximum likelihood loss is overall the main driving force for the convergence of the weights during training.

To carry out the weight optimisation, we employed a standard stochastic gradient descent approach, namely the adaptive learning rate, momentum-based Adam \citep[adaptive moment,][]{Kingma2018} optimiser (with parameters $\beta_1 = \beta_2 = 0.8$). 
We initialised the Adam optimiser with a learning rate of $l_{\mathrm{init}} = 0.001$ (in Adam this corresponds to a scaling factor applied to the adaptive step size in the weight updates) and we made use of the common L2 weight regularisation formalism with $\lambda = 1.0 \times 10^{-5}$ in the optimisation process.
The latter reduces the risk of overfitting to the training data by penalizing large weights in the network, which encourages convergence to a more generalisable solution \citep{Goodfellow2016_DeepLearning}. 
In addition, we adopted a decaying learning rate schedule, reducing $l$ by a factor of $\gamma = 0.95$ after every training epoch and we trained the cINN for 123 epochs (stopping early once the training loss converged) in total, processing $128$ batches of size $1024$ in every epoch (i.e.~30\% of the total training data). 
We note that we did not loop over the entirety of our training data in every epoch here for the purpose of reducing the overall training time. 
Training a single cINN took about 7.5 h when using a NVIDIA RTX 6000 Ada graphics card for GPU acceleration. 

\subsection{Hyperparameter search}
\label{sec:hyperparameters}
To derive the hyperparameters of our cINN architecture outlined in Sects.~\ref{sec:training_strategy} and \ref{sec:architecture_details} above, we employed an optimisation approach called Hyperband \citep{Li2018_Hyperband}. It uses early stopping and adaptive resource allocation to provide a more efficient framework for random grid hyperparameter searches. In the Hyperband set-up, a maximum amount of training epochs per one network configuration is specified as a budget. Based on this maximum budget, Hyperband first generates differently sized groups of randomly generated hyperparameter configurations for the network. Based on the group size, Hyperband then employs a more or less aggressive pruning tactic during the hyperparameter search process (the larger the group, the more aggressive is the pruning). This pruning strategy determines for how many epochs at a time the different network configurations are allowed to train before the least promising set-ups (based on the validation loss) are discarded. Specifically, the largest group trains only for a few epochs at a time before pruning, whereas smaller groups are allowed to train longer. In each group, this discarding procedure proceeds iteratively until only one network configuration is allowed to train for the specified maximum training epoch budget. In the final step, the 'winning' configurations in each group are then ranked by their validation loss. This hyperparameter search strategy tries to strike a balance between only training configurations until the end that appear promising early (i.e.~converge fast) and allowing for some slower converging models to be trained as well, which could potentially converge to a better solution late in training. For more details on the method, we refer to \cite{Li2018_Hyperband}.

\subsection{Point estimates and sample rejection}
\label{sec:point_estimation}

\begin{figure*}
    \centering
    \includegraphics[width=0.49\linewidth]{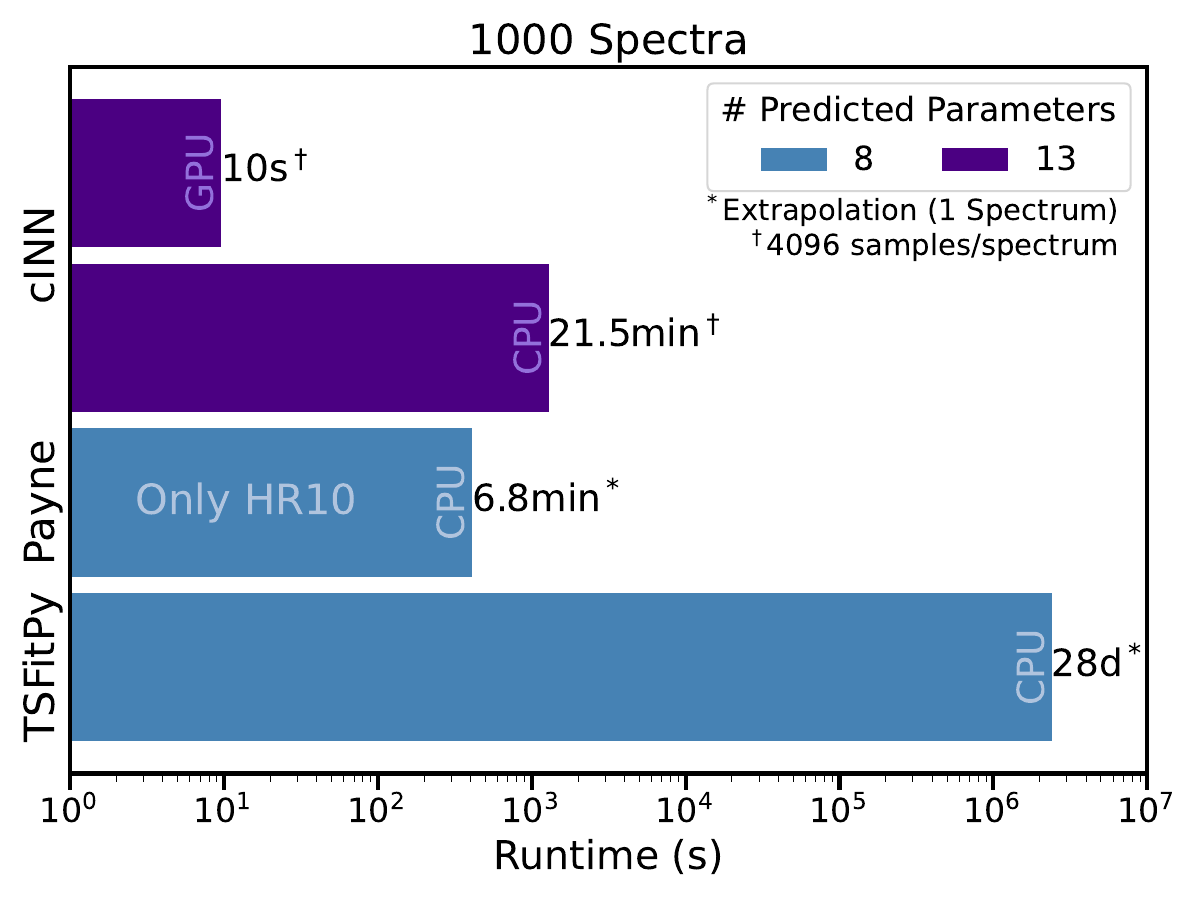}
    \includegraphics[width=0.49\linewidth]{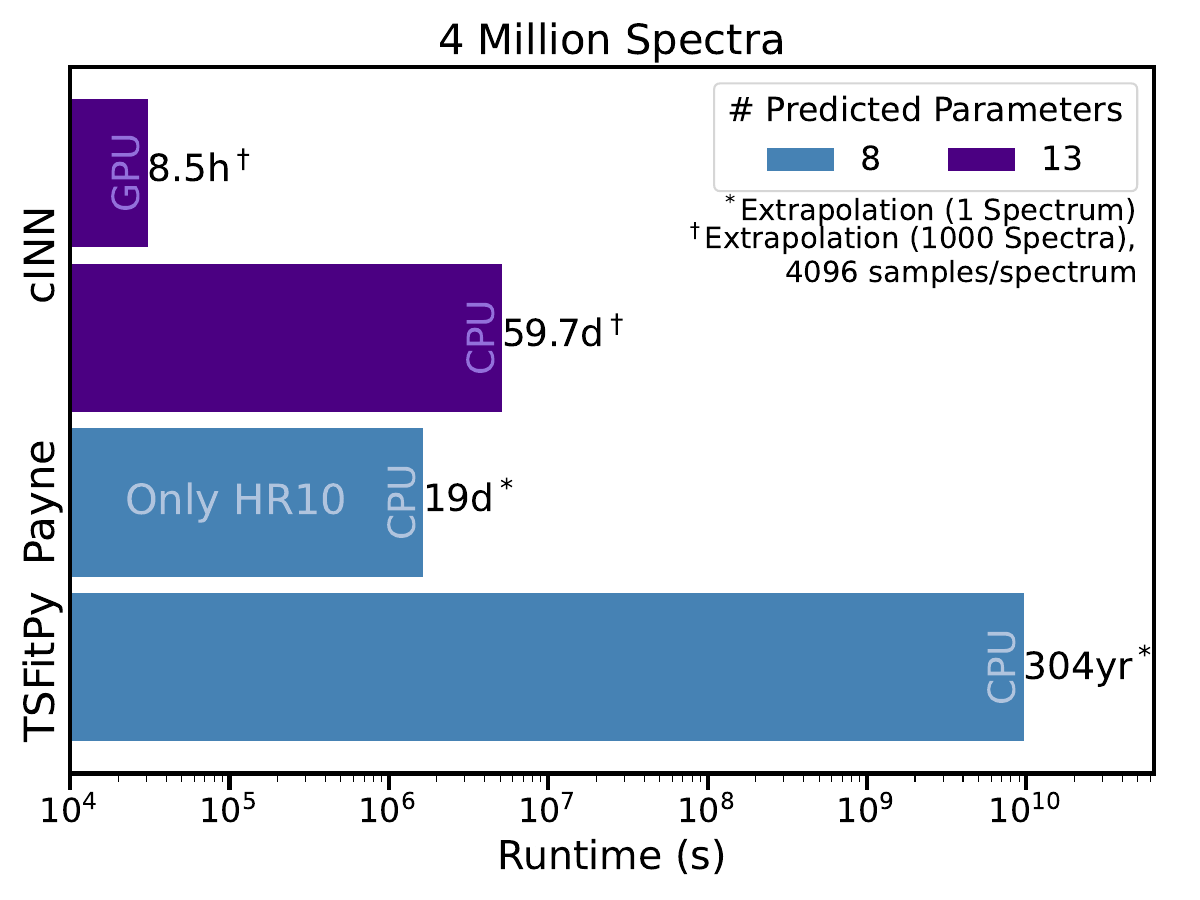}
    \caption{Runtime comparison between the proposed cINN approach to the previously established methods TSFitPy and the Payne. The performance of the cINN was measured for the case of generating 4096 posteriors samples per spectrum, using one core of an AMD EPYC 9554P (3.7 GHz) CPU and an NVIDIA RTX 6000 Ada GPU, respectively. The total cINN runtime is a combination of posterior generation and subsequent point estimation, but the latter becomes negligible quickly. We note that the runtimes in the right panel are extrapolated from the 1000 spectra (cINN) and 1 spectrum (TSFitPy and Payne) results, respectively. The Payne approach shown here operates on a more limited input parameter space, denoted as HR10, which covers the wavelength range from $534-562$ nm and is based on the set-up of the GIRAFFE spectrograph.}
    \label{fig:timing_comparison}
\end{figure*}

As outlined in Sect.~\ref{sec:cinn_intro}, our approach returns an estimate of the full posterior distribution of the target parameters given the input observations. However, for comparison with e.g.~ground truth values in the subsequent performance analysis, it is often instructive to provide point estimates based on the predicted posterior distributions. In the following, we describe how we used the maximum a posteriori (MAP) estimate for this purpose, namely, the most likely combination of target parameters as determined by the posterior distribution. To find this most likely solution, we can use the cINN itself, as it  not only generates samples of the posterior distributions, but also estimates the probability density of each generated sample. For each query observation, the MAP estimates are then identified as the posterior sample with the highest estimated probability density in the full 13 dimensional target parameter space of our model. 
In the following analysis presented below, we always generated 4096 posterior samples for each query observation, from which the MAP estimates are then determined. In our experiments, this number of samples has proven reliable enough to produce robust MAP estimates; thus, increasing the number of generated samples beyond this value did not markedly change the returned MAP values. 

It is possible that some of the generated posterior samples fall outside of the boundaries of the training data. There are two main reasons for this. Firstly, the cINN always returns smooth posterior distributions. Consequently, it does not handle hard boundaries very well. If a prediction is close to the edges of the trained parameter space, the predicted posterior can, therefore, extend slightly beyond the formal limits of the training data. In the second scenario, the input observation is either just outside the space spanned by the training examples or vastly different altogether. In the former case, some slight extrapolation can be expected, resulting in a fraction of posterior samples to fall outside the training boundaries, whereas in the latter the cINN will likely have to extrapolate significantly. As with most machine learning approaches, there is no guarantee that the prediction is still sensible, if the model has to extrapolate far from its training boundaries \citep{Bishop2006_PatternRecognitionAndMachineLearning, Goodfellow2016_DeepLearning}. To mitigate this potential issue, we included a sample rejection approach in our analysis of the predicted posteriors (e.g.~when we determine the MAP). Specifically, we rejected all posterior samples that fell outside the target parameter boundaries of the training data (for the limits see Fig.~\ref{fig:ts_priors}). In principle, we could also use softer margins at the training boundaries to allow for some minor extrapolation, but for the analysis presented in this work, we decided to be strict. Using this rejection approach, we also introduced a 'reliability' metric. Specifically, for each query spectrum, we determined the fraction of rejected posterior samples, $f_\mathrm{reject}$ (out of the 4096 initially generated ones). If $f_\mathrm{reject}$ is large (i.e. for ~$f_\mathrm{reject} > 0.5$), the cINN extrapolated more often; thus, the prediction might have to be treated with caution.   

\begin{figure*}[h!]
    \centering
    \includegraphics[width=\linewidth]{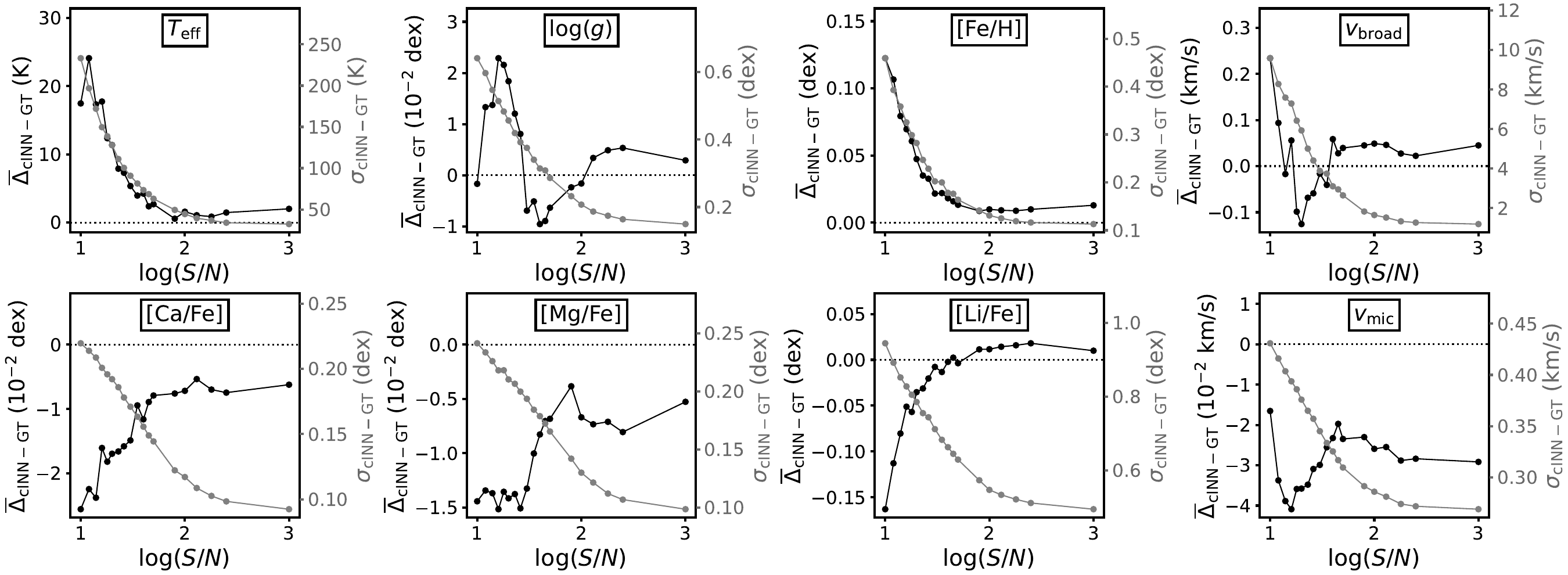}
    \caption{Summary of cINN performance on synthetic spectra as a function of S/N. In each panel, the mean residual $\overline{\Delta}_\mathrm{cINN-GT}$ is plotted in black on the left y-axis scale, while the standard deviation $\sigma_\mathrm{cINN-GT}$ is indicated in grey and plotted on the right y-axis scale. The black dotted line indicates, where $\overline{\Delta}_\mathrm{cINN-GT} = 0$ for reference. For more details, see Table~\ref{tab:me_summary} and Fig.~\ref{fig:Synth_ME_SIGMA_vs_SNR_DETAILED} in the appendix.}
    \label{fig:Synth_ME_SIGMA_vs_SNR}
\end{figure*}

\subsection{INNs versus other generative approaches}
As illustrated in the introduction, INNs are not the only generative deep learning approach that have seen usage in the context of spectra analysis.
Conceptionally, models such as variational autoencoders \citep[VAEs;][]{Kingma2013, Rezende2014} or generative adversarial networks \citep[GANs;][]{GoodfellowNIPS2014}, or (more specifically) their conditional versions (cVAEs; \cite{Sohn2015}, or cGANs; \cite{Mirza2014, Isola2017}), could be applied to solve inverse problems in a similar fashion to INNs \citep{Ardizzone2019a}. 
While concrete performance will naturally vary depending on the individual architectures used, INNs or NFs generally offer some advantages. 
One central difference of NFs to VAEs or GANs is that they can not only efficiently generate samples of a target distribution, but, at the same time, also compute an exact density estimate for the generated samples \citep{Kobyzev2021, BondTaylor2022}. As outlined in Sect.~\ref{sec:point_estimation}, we used the latter capability, for instance, to derive the MAP estimates from the generated posteriors.
In addition, \cite{Ardizzone2019a} demonstrated that INNs tend to return better calibrated posterior distributions and more accurate point estimates than a comparable cVAE approach. 
The posterior distributions predicted by INNs also default to the prior distribution of a certain target parameter if the conditioning input (i.e.~input observation) does not contain enough information to constrain the parameter, whereas cVAEs may return spurious distributions in this case \citep{Ardizzone2019a}. 
In general, as VAEs only offer limited posterior approximations, they are more likely to underestimate variance in the posteriors, to induce biases in their MAP estimates and to suffer from mode collapse \citep{Kobyzev2021, BondTaylor2022}. 
The latter issue is also often encountered in GANs \citep{Ardizzone2019b, Kobyzev2021, BondTaylor2022}. In contrast, if trained via maximum likelihood, INNs are extremely robust to mode collapse by construction, as it is heavily penalised in this training strategy \citep{Ardizzone2019b}. 
Furthermore, maximum likelihood training is very stable, contrary to the adversarial training regime in GANs, which is notoriously hard and unstable \citep{Ardizzone2019b, BondTaylor2022}. 
Lastly, \cite{Liu2020} also demonstrated that GANs struggle significantly with modelling low-dimensional density distributions in comparison to NFs.

\section{Results}
\label{sec:results}
In this section, we summarise the performance of our trained cINN in terms of runtime (Sect.~\ref{sec:results_runtime}), on synthetic stellar spectra (Sect.~\ref{sec:results_synthetic}) and on observed archival spectra of the Gaia-ESO/PLATO/4MOST benchmark stars (Sect~\ref{sec:results_benchmark}). We note that the following focuses on the predictive performance results for \teff, \logg, \vmic, \vbroad, \feh, [Ca/Fe], [Mg/Fe], and [Li/Fe]. Even though our NLTE cINN also (jointly) predicts the abundances [C/Fe], [Mn/Fe], [N/Fe], [Si/Fe], and [Ti/Fe], we do not provide results for these chemical elements in this paper. A validation of the other elemental abundances will be a subject of a follow-up work by our group.

\subsection{Network efficiency}
\label{sec:results_runtime}
Figure~\ref{fig:timing_comparison} provides a summary of our cINN runtime, consisting of posterior generation plus subsequent point estimation according to Sect.~\ref{sec:point_estimation}, for example datasets of one thousand spectra and four million spectra, respectively. As demonstrated, the cINN approach is highly efficient and takes only 21.5~min on a single CPU core (AMD EPYC 9554P $@$3.7~GHz) or 10~s on a GPU (NVIDIA RTX 6000 Ada) to process 1000 query spectra without leveraging any additional parallelisation. Based on this, the cINN could process the output of, for instance, the 4MIDABLE-HR survey, expected to obtain more than four million spectra with the 4MOST facility\footnote{We note that this is the estimate of the number of spectra, whereas the number of stars to be observed is lower. For many stars, multi-epoch spectroscopic data will be available.}, in less than half a day when using GPU acceleration on a single GPU (when generating 4096 posterior samples per spectrum; i.e.~more than $10^{10}$ posterior samples in total). For context, Fig.~\ref{fig:timing_comparison} also shows a comparison of the cINN runtime to the state-of-the-art classical fitting method TSFitPy, which is also our primary benchmark for the elemental abundances on the real spectra in Sect.~\ref{sec:results_benchmark}. This highlights the great advantage of deep learning approaches, such as the cINN, as it would require 28 days (when determining only eight parameters on a single CPU) to process just 1000 spectra (or more than 300 years for four million spectra), for example. For reference, Fig.~\ref{fig:timing_comparison}, also shows the runtime of a different machine learning method, namely the neural network approach known as Payne. We note that the Payne set-up used here operates on a smaller input space (denoted as HR10) that covers a wavelength range of $534$ - $562$ nm and is based on the set-up of the GIRAFFE spectrograph. This comparison demonstrates that the cINN runtime is on the same order of magnitude as the Payne, while, at the same time, accounting for a much larger input space, predicting five parameters more, and returning full posterior distributions in addition to a point estimate.

\subsection{Tests on synthetic spectra}
\label{sec:results_synthetic}

\begin{figure*}[ht!]
    \centering
    \includegraphics[width=\linewidth]{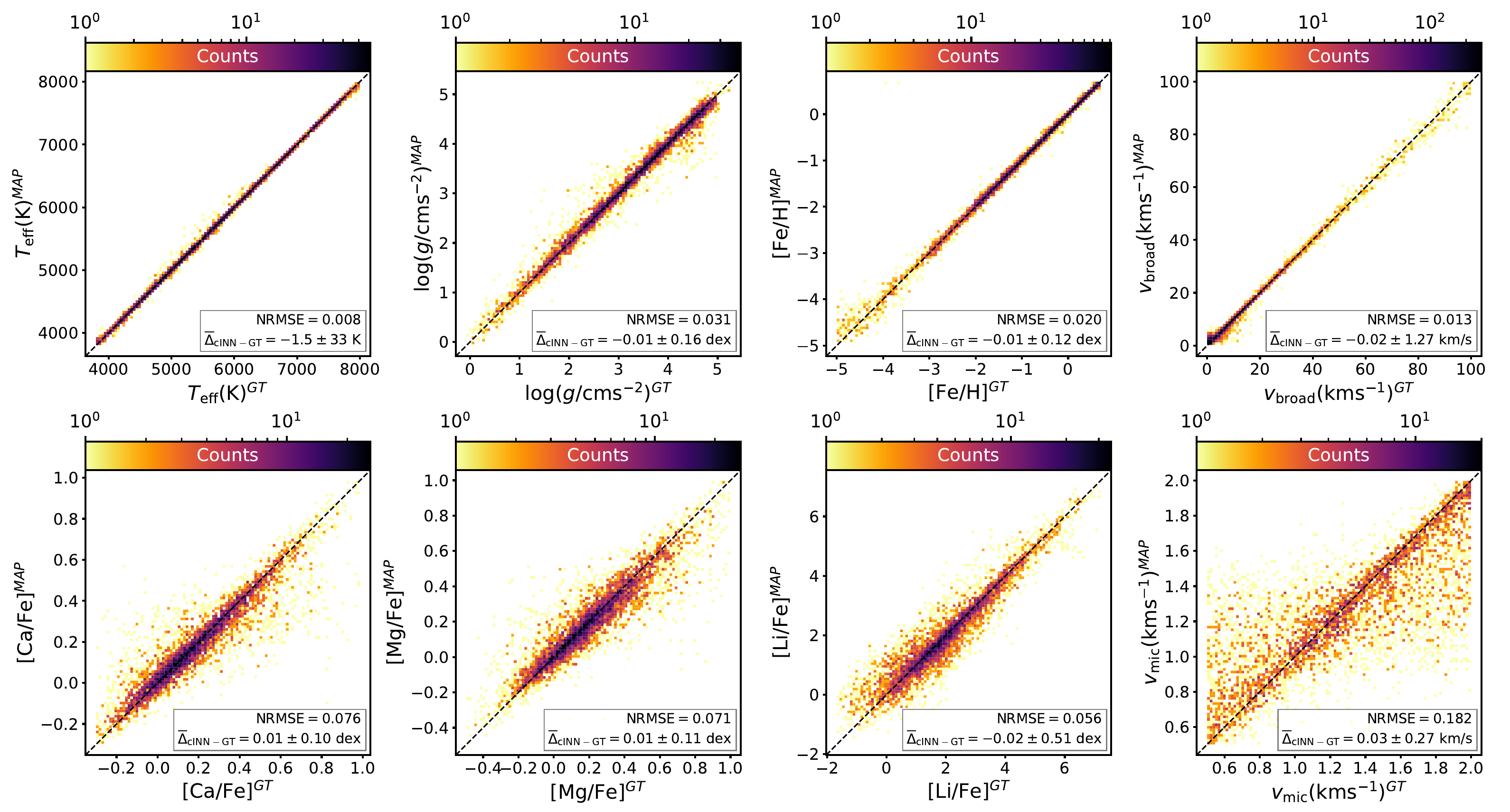}
    \caption{2D histograms that compare the cINN MAP estimates to the corresponding ground truth for 5910 synthetic spectra with $\mathrm{S/N} = 250$ held-out in the test set. The dotted black guide line indicates a perfect one-to-one match.}
    \label{fig:map_vs_gt_snr250}
\end{figure*}

After training the cINN, we evaluated the predictive performance on the $N=5910$ dedicated, held-out synthetic test spectra. In this, we computed two metrics based on the MAP estimates of the cINN $X_\mathrm{MAP}$ with respect to the ground truth, $X_\mathrm{GT}$, to quantify the performance. The first are the mean, $\overline{\Delta}_\mathrm{cINN-GT}$,
\begin{equation}
    \label{eq:mae}
    \overline{\Delta}_\mathrm{cINN-GT}(X) = \frac{1}{N}\sum_{i=1}^{N} X_{i, \mathrm{MAP}} - X_{i, \mathrm{GT}},
\end{equation}
and the standard deviation $\sigma_\mathrm{cINN-GT}$,
\begin{equation}
    \label{eq:sd_ae}
    \sigma_\mathrm{cINN-GT}(X) = \sqrt{\frac{\sum_{i=1}^{N}\left| (X_{i, \mathrm{MAP}} - X_{i, \mathrm{GT}}) -\overline{\Delta}_\mathrm{cINN-GT}(X)\right|^2}{N-1}},
\end{equation}
of the residuals, used to gauge bias in the cINN MAP estimates and the average error, respectively. The second is the normalised root mean squared error (NRMSE), expressed as
\begin{equation}
    \mathrm{NRMSE} = \frac{\sqrt{\frac{1}{N}\sum_{i=1}^{N} (X_{i, \mathrm{MAP}} - X_{i, \mathrm{GT}})^2}}{\hat{X}_\mathrm{max} - \hat{X}_\mathrm{min}},
\end{equation}
where $\hat{X}_\mathrm{max}$ and $\hat{X}_\mathrm{min}$ refer to the upper and lower boundaries of target parameter $X$ in the training data, that serves to compare the predictive performance between different target parameters. In Fig.~\ref{fig:Synth_ME_SIGMA_vs_SNR}, we provide a summary of the cINN predictive performance in terms of $\overline{\Delta}_\mathrm{cINN-GT}$ and $\sigma_\mathrm{cINN-GT}$ across all S/N that the network is trained on (see also Tables~\ref{tab:me_summary} and \ref{tab:nrmse_summary} in the appendix). 
As we can see, the network performance depends notably on the quality of the input spectra as $\sigma_\mathrm{cINN-GT}$ (and the NRMSE) increases across all target parameters the more noisy the input observations become. The bias $\overline{\Delta}_\mathrm{cINN-GT}$ (although overall small) likewise tends to grow larger (in magnitude), with trends towards overestimation for \teff~and \feh, tendencies towards underestimation for [Ca/Fe], [Mg/Fe] and [Li/Fe], and mixed behaviour for \logg, \vbroad~and \vmic~as S/N decreases (for more details see Appendix~\ref{app:synth_test} and Fig.~\ref{fig:Synth_ME_SIGMA_vs_SNR_DETAILED}).  
This is accompanied by a widening of the predicted posterior distributions with decreasing S/N, as illustrated by the median width of the 68\% confidence interval $\Delta_\mathrm{68\%}$ (of each 1D component of the predicted posterior) in Fig.~\ref{fig:u68_vs_snr} in the appendix. Here, in particular, the abundances become quite uncertain at low S/N, exhibiting particularly wide posteriors.
This result demonstrates that the cINN, on the one hand, correctly internalised that an increased level of white noise (which is assumed, but may not obviously be true in observed data) in the input fluxes warrants a greater uncertainty in the predicted parameters. 
On the other hand, it also indicates that the stellar parameters and abundances become harder to constrain for the cINN at low S/N, as evidenced by the increased $\sigma_\mathrm{cINN-GT}$ (and NRMSE) of the MAP point estimates. Specifically, $\overline{\Delta}_\mathrm{cINN-GT}\pm\sigma_\mathrm{cINN-GT}$ increases from $2\pm32$ K, $0.003\pm0.15$ dex, $0.01\pm0.11$, $0.05\pm1.19$ km\,s$^{-1}$, $-0.03\pm0.27$ km\,s$^{-1}$, $-0.006\pm0.09$ dex, $-0.005\pm0.1$ dex and $0.01\pm0.49$ dex at $\mathrm{S/N} = 1000\,$\AA$^{-1}$ for \teff, \logg, \feh, \vbroad, \vmic, [Ca/Fe], [Mg/Fe] and [Li/Fe], respectively, to $18\pm233$ K, $-0.002\pm0.64$ dex, $0.12\pm0.46$ dex, $0.23\pm9.59$ km\,s$^{-1}$, $-0.017\pm0.43$ km\,s$^{-1}$, $-0.026\pm0.22$ dex, $-0.014\pm0.24$ dex and $-0.16\pm0.94$ dex at $\mathrm{S/N} = 10\,$\AA$^{-1}$.   
As demonstrated in Fig.~\ref{fig:Synth_ME_SIGMA_vs_SNR} (see also Fig. \ref{fig:u68_vs_snr} in the appendix), our final cINN performs best in the S/N range between 80 and 1000 \AA$^{-1}$, maintaining a comparable performance. Below $\mathrm{S/N} = 80\,$\AA$^{-1}$, however, both intrinsic uncertainties (i.e.~posterior widths) and average errors of the prediction start to increase significantly. 
We note that for the analysis of low S/N observations, a more specialised cINN, trained only on low S/N data, could potentially perform better than the general purpose model presented here.

Next, we discuss the results obtained with the NLTE cINN on synthetic spectra at $\mathrm{S/N} = 250\,$\AA$^{-1}$ more in detail, as this is the expected S/N of the high-resolution spectra of stars to be observed in the 4MOST 4MIDABLE-HR program.
Figure~\ref{fig:map_vs_gt_snr250} shows a 2D histogram comparing the cINN MAP estimates on the synthetic test spectra to the ground truth. 
At this S/N, we find that our NLTE cINN recovers atmospheric parameters and abundances of stars with $\overline{\Delta}_\mathrm{cINN-GT}\pm\sigma_\mathrm{cINN-GT}$ values of $1.46\pm33$ K, $0.005\pm0.16$ dex, $0.01\pm0.12$ dex, $-0.007 \pm 0.1$ dex, $-0.008\pm0.11$ dex, $0.02\pm0.51$ dex, $0.02 \pm 1.27$ km\,s$^{-1}$ and $-0.03 \pm 0.27$ km\,s$^{-1}$ for \teff, \logg, \feh, [Ca/Fe], [Mg/Fe], [Li/Fe], \vbroad~and \vmic, respectively. Looking at the NRMSEs, we find that \teff~is comparatively recovered the best with an NRMSE of 0.008, whereas \vmic~appeared the hardest to constrain with an NRMSE of 0.182.  

Analogously to the analysis of the posterior width as a function of S/N, we also investigated how the cINN performance behaves in dependence of the primary stellar parameters, \teff, \logg~and \feh~for the fixed $\mathrm{S/N} = 250\,$\AA$^{-1}$.
This analysis revealed that the chemical composition, and in particular [Fe/H], is the most critical parameter that influences the quality of the results.
Specifically, for $[\mathrm{Fe/H}] \leq -2.5$, the predicted posteriors for the abundances become so wide that the cINN returns the entire training parameter range as a possible possible solution (see Fig.~\ref{fig:full_posterior_vs_feh} for an illustration of the full posterior width as a function of \feh). This increase in the intrinsic uncertainty of the cINN prediction is also reflected by increased $\sigma_\mathrm{cINN-GT}$ values, rising, for example, by a factor of 2 for [Mg/Fe] or [Ca/Fe] when compared to synthetic spectra with $\mathrm{[Fe/H]} > -2.5$. 
One potential explanation for this behaviour is the large dynamical range of chemical abundances spanning five orders of magnitude in all key elements, such as C, Fe, Mg, Ti. This leads to significant variations in atomic and molecular opacity over the parameter space of FGKM-type stars, being especially critical for red giants with their lower surface temperatures favouring formation of molecules \citep{Plez2012, Eitner2025}.
Apart from that, the grid of available synthetic stellar spectra is rather sparse for $[\mathrm{Fe/H}] \lesssim -2$. This regime is typically associated with very metal-poor (VMP) and extremely metal-poor (EMP) stars \citep{Christlieb2005}. Due to this relative scarcity of training examples, the cINN cannot achieve the same precision in this (VMP) metallicity regime, which was also demonstrated in other studies employing machine learning \citep[see Table~1 in][]{Kovalev2019}. We note that there are efforts within the 4MOST Infrastructure Group 7 (IWG7) to compute dedicated spectral grids in this domain, and therefore we reserve the optimisation of our cINN approach in the VMP and EMP regime to a follow-up study.

Finally, we explored the decision-making process of the trained cINN by computing saliency maps for all our target parameters. Saliency describes the average network internal gradients of each target parameter with respect to each input feature $\partial X /\partial f_\lambda$ (where $f_\lambda$ is the input flux for a given wavelength bin). 
Looking, in particular, at the resulting saliency maps of the predicted abundances (see Figs.~\ref{fig:saliency1} and \ref{fig:saliency2} in the appendix for an illustration), we found that the most significant peaks in the saliency (i.e.~the features with the highest importance) almost always coincide with absorption lines of the corresponding element (see e.g.~the [Ca/Fe], [Mg/Fe] and [Li/Fe] panels in Fig.~\ref{fig:saliency2}). This indicates that the trained cINN has learned to pay attention to parts of the spectrum, i.e.~the spectral lines of different elements, that also play an important part in classical analysis techniques (albeit not necessarily in the same exact combination). 
This confirms that the decision making process of the trained network has picked up on some of the physics used in the computation of the synthetic stellar spectral grids. 
This is not unexpected, because the cINN is essentially a function - the result of an optimisation problem (here, training the cINN) - and by definition mathematically constructed such that the input data are reproduced as closely as possible given the degeneracies associated with the multi-dimensionality.

\subsection{Test on benchmark spectra}
\label{sec:results_benchmark}

\begin{figure*}[ht!]
    \centering
    \includegraphics[width=0.95\linewidth]{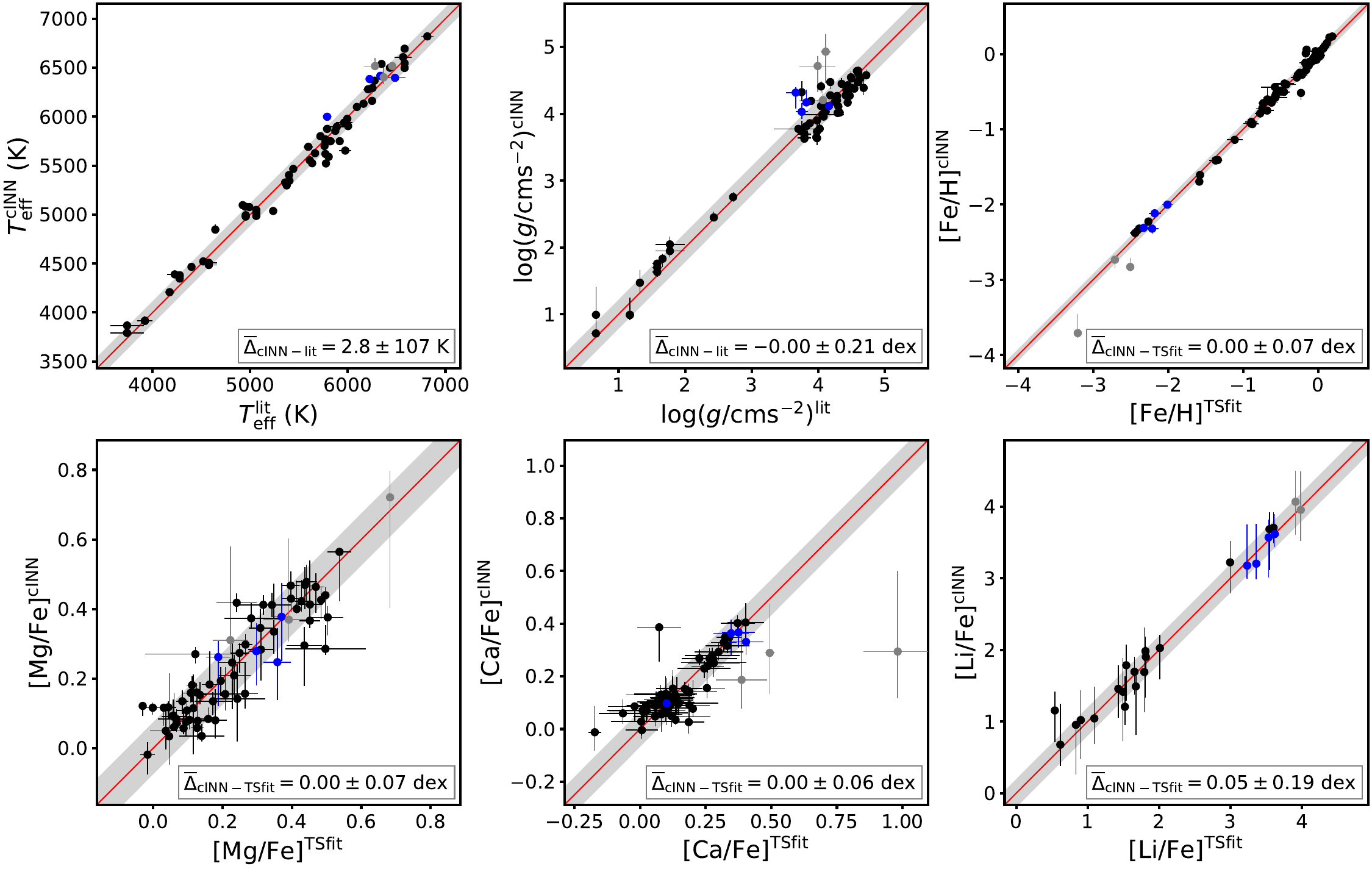}
    \caption{Comparison of cINN predictions after correction for systematic biases to the literature stellar parameters ($X^\mathrm{lit}$) and abundance fits with Turbospectrum ($X^\mathrm{TSfit}$) for the real benchmark spectra listed in Table~\ref{tab:benchmark_params}. We note that some stars have multiple spectra in our benchmark set. For some spectra, there are no uncertainties listed in the literature for $T_\mathrm{eff}$ and $\log(g)$, so no x-error bars are displayed in these cases (see Table~\ref{tab:benchmark_params}). The $T_\mathrm{eff}$ uncertainties are often too small to be visible in the top left panel. The y-error bars indicate the 1D 68\% confidence intervals determined from the predicted posterior distributions. Stars indicated in grey have $\mathrm{[Fe/H]} \leq -2.5$ and were not considered in the fit for the bias correction (but are corrected here). The grey shaded area around the 1-to-1 correlation represents the 1$\sigma$ standard deviation of the residuals of the MAP estimate with respect to the reference value $X^\mathrm{cINN, MAP} - X^\mathrm{lit/TSfit}$. The stars highlighted in blue correspond to the example stars shown in Figs.~\ref{fig:resim_Mg_Ca_Li} and \ref{fig:resim_greenredwindow}.}
    \label{fig:benchmark_1to1_corr}
\end{figure*}

After verifying the network performance on the synthetic test spectra, we applied the cINN to the set of archival spectra for the benchmark stars, as outlined in Sect.~\ref{sec:data_benchmark}. Stellar parameters, in particular \teff~and \logg, for these targets were assembled from the literature (see Table~\ref{tab:benchmark_params}). For the Gaia-ESO benchmark stars, \teff~and \logg~were obtained using methods that are almost independent of spectroscopy \citep{Heiter2015,Jofre2014}. In particular the \teff~values were constrained using interferometric angular diameters, whereas the \logg~values were obtained either on the basis of evolutionary phase considerations (models of stellar interior) or from global asteroseismic parameters. For the stars observed with the FOCES instrument (Sect. 2.2), however, no such constraints are available, therefore their estimates rely on spectroscopy. 

Since [X/H] abundance ratios are very sensitive to the physics of stellar spectral models and of the spectroscopic analysis approach \citep{Bergemann2012a, Lind2012, Bergemann2017, Bergemann2019, Amarsi2019, Amarsi2024}, we adopted our own abundances for these benchmark stars. Here we used the TSFitPy code (see Sect.~\ref{sec:data}) to determine NLTE values of [Fe/H], [Li/Fe], [Mg/Fe], and [Ca/Fe] for each star in Table~\ref{tab:benchmark_params} directly. The diagnostic lines are provided in Table~\ref{tab:tsfitpy_lines} (appendix). In these abundance calculations, the values of \teff, \logg, \vbroad~and \vmic were set to the values of the cINN MAP estimates. The results of our TSFitPy calculations are provided in  Table~\ref{tab:benchmark_predictions}. For some spectra, it was not possible to determine some elemental abundances, such as [Li/Fe], because of the weakness of the spectral features.

When comparing our initial cINN predictions for stellar parameters and abundances with the afore-mentioned estimates, we identified a small bias in the predictions, which correlates with \teff~(see Table~\ref{tab:benchmark_predictions_raw} for the raw cINN predictions). This bias is also known from previous studies with the Payne approach \citep{Gent2022}, which requires the use of line masks in the form of a wavelength-dependent noise model, or alternatively more sophisticated Bayesian-supported noise models to improve the solutions \citep{Bestenlehner2024}. Here, we introduced a simple linear correction (see Appendix~\ref{app:Bias_corr} and Table~\ref{tab:calib_coeff} for details), which removes the underlying \teff-dependent bias in our parameter and abundance estimates. As a result of applying this correction, we were able to achieve stable and consistent results for the entire parameter space of our observed FGKM-type stars.

In Fig.~\ref{fig:benchmark_1to1_corr}, we show our resulting stellar parameters and chemical abundances obtained with the NLTE cINN (y-axis) in comparison to those obtained with TsFitPy and from literature sources (x-axis). Our final values are also provided in Table~\ref{tab:benchmark_predictions}. For all elements (\feh, [Ca/Fe], [Mg/Fe] and [Li/Fe]), the agreement between independent estimates is very good and it is typically within the combined uncertainties of both values. The typical errors ($\sigma_\mathrm{cINN-lit/TSfit}$) of our predictions are $107$ K for \teff, $0.21$ dex for \logg, $0.07$ dex for \feh, $0.07$ dex for [Mg/Fe], $0.06$ dex for [Ca/Fe], and $0.19$ dex for [Li/Fe], respectively.
It is important to note here, that our correction currently extends only down to $[\mathrm{Fe/H} \approx -2.5$ (for the reasons outlined in the previous section) and is limited to an interval of 0.5 to 3.9 in [Li/Fe]. The predictions in Table~\ref{tab:benchmark_predictions} for the few very metal poor stars and where $[\mathrm{Li/Fe}]^\mathrm{cINN} < 0$ may, therefore, need to be treated with caution. 
In a greater context, the performance of our cINN does not only allow for a robust analysis of observed spectra, as indicated by the successful recovery of the Turbospectrum NLTE abundances, but should also be sufficiently accurate to facilitate studies that aim to chemically distinguish Galactic populations, such as the thin and thick disc, as demonstrated for example by \cite{Gent2024}.

\begin{figure*}[h!]
    \centering
    \includegraphics[width=\linewidth]{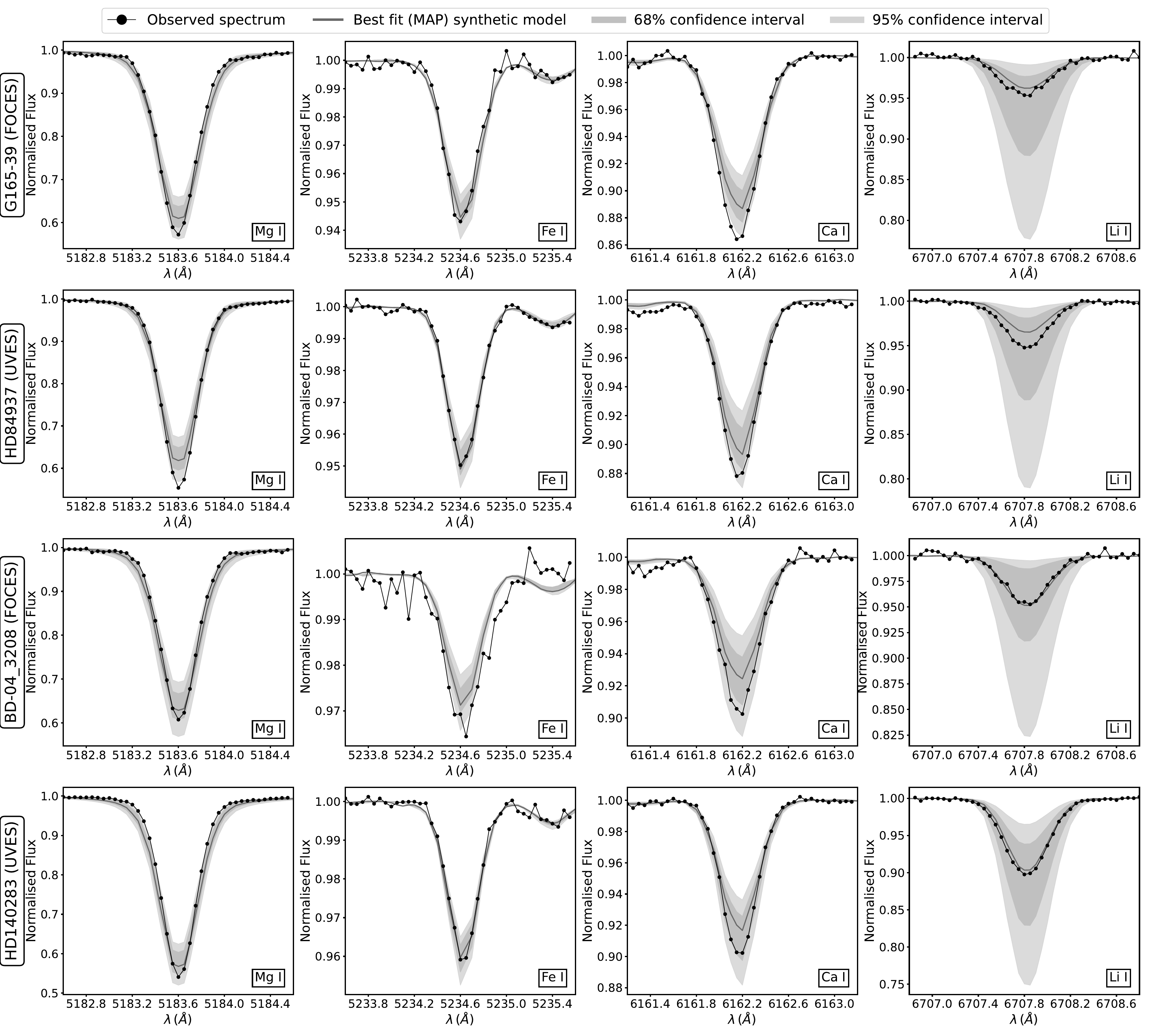}
    \caption{Comparison of the 4MOST-ified input spectra to model spectra corresponding to the cINN MAP estimates for four low-metallicity stars, focusing on prominent absorption lines of Mg, Ca, Li and Fe. We note that the re-simulation employed all 13 stellar parameters predicted by the cINN, but used the bias correction from Sect.~\ref{sec:results_benchmark} only for the abundances of Ca, Mg, Li, and Fe (see also Table~\ref{tab:calib_coeff}).}
    \label{fig:resim_Mg_Ca_Li}
\end{figure*}

As a final, qualitative evaluation of the NLTE cINN performance on the real data, we used Turbospectrum \citep{Gerber2023, Storm2023} to compute theoretical spectra that correspond to the cINN estimates of stellar parameters and abundances. Our model spectra are then compared to the observed wavelength-dependent datasets.
We note that for this analysis, we used all 13 output parameters of the cINN, but only applied the bias correction to the four abundances (i.e.~\feh, [Mg/Fe], [Ca/Fe], and [Li/Fe]) it is available for.
The reason for this is that there is no one-to-one mapping between stellar parameters and the cINN functional behaviour, and stellar parameters have a global non-linear impact on the spectra, also affecting the shape of the continuum. For abundances it is, however, appropriate to do this, because the diagnostic elements (Mg I, Fe I, Li I, Ca I) are trace (minority) species in FGKM-type stellar atmospheres \citep{Bergemann2014} and have no impact on the optical continuum. Therefore, by applying abundance corrections we can be confident that only the fluxes in the lines of these specific elements are affected.
Furthermore, to visualise the effect of the uncertainties in the predicted abundances of [Fe/H], [Mg/Fe], [Ca/Fe], and [Li/Fe], we kept all nine other predicted parameters fixed to the MAP estimate and computed four additional models using the 2.5\%, 16\%, 84\%, and 97.5\% quantiles of the abundances to approximate the 68\% and 95\% confidence intervals of the re-simulated flux.

In Fig.~\ref{fig:resim_Mg_Ca_Li}, we show the observed spectra of four stars and overplot the simulated spectra computed for the cINN parameter predictions. We in particular highlight the diagnostic absorption lines of Mg I, Ca I, Li I, and Fe I. We refer to  Fig.~\ref{fig:resim_greenredwindow} in the appendix, where the spectra are presented in full. Generally, we find that the model spectra computed using the cINN estimates fit the observed data well. Averaged over all ($72 \times 25\,197$) wavelength bins of our benchmark set, the fluxes of the re-simulated spectra (corresponding to the MAP estimates) and the input observations match to within $0.69_{-0.52}^{+2.12}$ \%. 

\section{Summary and outlook}
\label{sec:summary}

In this paper, we introduce a new, simulation-based, machine-learning method that employs a conditional invertible neural network (cINN) to determine astrophysical parameters of stars from observed stellar spectra. The parameters include effective temperature, \teff, surface gravity, \logg, metallicity,~\feh, broadening, \vbroad, microturbulent velocity, \vmic~and the elemental abundance ratios, [Ca/Fe], [Mg/Fe], and [Li/Fe]. The framework presented can be applied to data with the spectral characteristics (resolving power, noise, wavelength coverage) of the 4MOST high-resolution ($\mathrm{R} = 20\,000$, wavelength coverage: $5160-5730$\,\AA, $6100-6790$\,\AA) spectrograph \citep{DeJong2019_4MOST}.

Our cINN was trained on a grid of $29\,548$ synthetic stellar spectra, computed in NLTE and using MARCS model atmospheres \citep{Gustafsson2008} with the Turbospectrum code \citep{Gerber2023, Storm2023}. We further augmented this base model grid by computing a version of each spectrum at 19 different S/N values, totalling $561\,412$ synthetic spectra. The latter were then split 75\%, 20\%, and 5\% into the dedicated training, test and validation sub-sets, respectively. 
The cINN is part of the normalising flow family and with that returns predictions of the full posterior distribution of the target stellar parameters and elemental abundances.
This allows the cINN to both provide an internally consistent uncertainty estimate and to highlight potential degeneracies in the inverse problem, while remaining highly efficient at inference time. The cINN presented in this study can evaluate 1000 high-resolution 4MOST spectra within a few 10s of seconds when making use of GPU acceleration (on an NVIDIA RTX 6000 Ada), which means it has the capability to process the entire output of a large 4MOST survey, such as e.g.~4MIDABLE-HR (over four  million spectra), in less than half a day.

We first tested the performance of our trained cINN on held-out synthetic test spectra, where we found that our NLTE cINN has the capacity to recover stellar parameters and abundances with average errors $\sigma$ (i.e.~standard deviation of the residuals) of $33$ K for \teff, $0.16$ dex for \logg, $0.12$ dex for \feh, $1.27$ km\,s$^{-1}$ for \vbroad, $0.27$ km\,s$^{-1}$ for \vmic, $0.1$ dex for [Ca/Fe], $0.11$ dex for [Mg/Fe], and $0.51$ dex for [Li/Fe], respectively, at an average $\mathrm{S/N} = 250\,$\AA$^{-1}$ of the input spectra, for example.
A more detailed analysis revealed that our NLTE cINN performs the best for stars with $[\mathrm{Fe/H}] > -2.5$ and spectra with $\mathrm{S/N} > 80\,$\AA$^{-1}$. Outside of this regime, we observed a significant increase in the intrinsic uncertainties of the cINN predictions (i.e.~width of the posteriors) that is accompanied by an increase in average errors $\sigma$. Here, we found that the \feh~dependency of the predictive performance can likely be attributed to a relative lack of training examples at very low metallicity in the available training data. The decrease in performance with S/N, on the other hand, could be explained by the general increase in complexity that comes with more noisy input data and the general purpose approach of the presented cINN that aimed to make the method applicable to a wide range of S/Ns from $10$ to $1000\,$\AA$^{-1}$; rather than specialising in~low-quality spectra alone, for instance.    

Next, we applied the NLTE cINN to a set of 72 real benchmark and bright star spectra, degraded to the resolution and wavelength coverage of 4MOST. We compared the predicted maximum a posteriori (MAP) estimates of the cINN to literature results for \teff~and \logg, and to \feh~and elemental abundances derived with classical evaluation techniques, using the TSFitPy code.
For this introductory paper, we focused here on the Ca, Mg, and Li abundances, reserving an analysis of the other elements to a follow-up study. 
In this comparison, we found a generally good agreement of the cINN estimates with the literature/TSFitPy results, but we identified a slight systematic offset as a function of the predicted \teff. To account for this apparent bias, we computed a linear correction (see Eq.~\eqref{eq:lin_corr} and Table~\ref{tab:calib_coeff}). When including this correction, we found an agreement of the cINN MAP estimates with the independent measures with $\sigma$ values of $107$ K for \teff, $0.21$ dex for \logg, $0.07$ dex for \feh, $0.07$ dex for [Mg/Fe], $0.06$ dex for [Ca/Fe], and $0.19$ dex for Li. 
This shows that a cINN-based approach is a promising avenue for the comprehensive and efficient NLTE analysis of large datasets to be obtained with the 4MOST facility in, for instance, the upcoming 4MIDABLE HR survey \citep{Bensby2019}. 

In a planned follow-up to this proof-of-concept study, we will extend our analysis to the remaining elemental abundances that are predicted by the network (i.e.~C, Mn, N, Si, and Ti) to verify their proper recovery. In addition, we plan to expand our suite of synthetic NLTE spectra with additional examples at very low metallicity to improve the cINN predictive performance in this regime. Further improvements can also be made in the S/N coverage by moving beyond the discrete set of options shown in this work to a continuous treatment of the S/N by using, for instance, a training strategy with dynamic data augmentation. Once we have realised these complementary analyses and improvements, we plan to release our model as a free tool to the astronomical community.

\begin{acknowledgements}
The team at Heidelberg University acknowledges financial support from the European Research Council via the ERC Synergy Grant ``ECOGAL'' (project ID 855130),  from the German Excellence Strategy via the Heidelberg Cluster  ``STRUCTURES'' (EXC 2181 - 390900948), and from the German Ministry for Economic Affairs and Climate Action in project ``MAINN'' (funding ID 50OO2206). They are also grateful for computing resources provided by the Ministry of Science, Research and the Arts (MWK) of the State of Baden-W\"{u}rttemberg through bwHPC and the German Science Foundation (DFG) through grants INST 35/1134-1 FUGG and 35/1597-1 FUGG, and also for data storage at SDS@hd funded through grants INST 35/1314-1 FUGG and INST 35/1503-1 FUGG. V.K. also acknowledges financial support by the Carl Zeiss Stiftung. R.A. and N.S. acknowledge support from the Lise Meitner grant from the Max Planck Society (grant PI: M. Bergemann) and from the European Research Council (ERC) under the European Union’s Horizon 2020 research and innovation programme (Grant agreement No. 949173, PI: M. Bergemann). G.G. acknowledges support by Deutsche Forschungs-gemeinschaft (DFG, German Research Foundation) – project-IDs: eBer-22-59652 (GU 2240/1-1 "Galactic Archaeology with Convolutional Neural-Networks: Realising the potential of Gaia and 4MOST").     
\end{acknowledgements}

\bibliographystyle{aa}
\bibliography{./bibtex}

\appendix

\section{Tests on synthetic data}
\label{app:synth_test}

\begin{table*}[ht!]
    \centering
    \small
    \caption{Overview of the cINN predictive performance in terms of the mean and standard deviation, $\overline{\Delta}_\mathrm{cINN-GT}\pm\sigma_\mathrm{cINN-GT}$, of the residuals of the target parameters with respect to the ground truth on synthetic test data for different S/N.}
    \label{tab:me_summary}
    \begin{tabular}{lrrrrrrrr}
    \toprule
    \multicolumn{1}{c}{S/N} & \multicolumn{1}{c}{$T_\mathrm{eff}$} & \multicolumn{1}{c}{\multirow{2}{*}{$\log\left(\frac{g}{\mathrm{cms}^{-2}}\right)$}} & \multicolumn{1}{c}{\multirow{2}{*}{$[\mathrm{Fe}/\mathrm{H}]$}} & \multicolumn{1}{c}{$v_\mathrm{broad}$} & \multicolumn{1}{c}{$v_\mathrm{mic}$} & \multicolumn{3}{c}{$[\mathrm{x}/\mathrm{Fe}]$} \\
      \multicolumn{1}{c}{(\AA$^{-1}$)} & \multicolumn{1}{c}{$(K)$} & & & \multicolumn{1}{c}{$(\mathrm{kms}^{-1})$} & \multicolumn{1}{c}{$(\mathrm{kms}^{-1})$} & \multicolumn{1}{c}{Ca} & \multicolumn{1}{c}{Mg} & \multicolumn{1}{c}{Li} \\
    \midrule
    1000 & $2.00\pm32$ & $0.003\pm0.15$ & $0.01\pm0.11$ & $0.05\pm1.19$ & $-0.029\pm0.27$ & $-0.006\pm0.09$ & $-0.005\pm0.10$ & $0.010\pm0.49$\\
    250 & $1.46\pm33$ & $0.005\pm0.16$ & $0.01\pm0.12$ & $0.02\pm1.27$ & $-0.028\pm0.27$ & $-0.007\pm0.10$ & $-0.008\pm0.11$ & $0.018\pm0.51$\\
    180 & $0.86\pm36$ & $0.005\pm0.17$ & $0.01\pm0.12$ & $0.03\pm1.33$ & $-0.029\pm0.27$ & $-0.007\pm0.10$ & $-0.007\pm0.11$ & $0.016\pm0.52$\\
    130 & $1.07\pm39$ & $0.003\pm0.19$ & $0.01\pm0.12$ & $0.05\pm1.52$ & $-0.025\pm0.28$ & $-0.005\pm0.11$ & $-0.007\pm0.12$ & $0.014\pm0.53$\\
    100 & $1.58\pm44$ & $-0.002\pm0.21$ & $0.01\pm0.13$ & $0.05\pm1.65$ & $-0.026\pm0.29$ & $-0.007\pm0.12$ & $-0.007\pm0.13$ & $0.012\pm0.55$\\
    80 & $0.56\pm49$ & $-0.002\pm0.23$ & $0.01\pm0.14$ & $0.04\pm1.83$ & $-0.023\pm0.29$ & $-0.008\pm0.12$ & $-0.004\pm0.14$ & $0.011\pm0.57$\\
    50 & $2.68\pm62$ & $-0.006\pm0.29$ & $0.01\pm0.16$ & $0.04\pm2.64$ & $-0.023\pm0.31$ & $-0.008\pm0.14$ & $-0.007\pm0.17$ & $-0.004\pm0.63$\\
    45 & $2.37\pm68$ & $-0.009\pm0.31$ & $0.02\pm0.18$ & $0.03\pm2.95$ & $-0.020\pm0.32$ & $-0.009\pm0.15$ & $-0.007\pm0.17$ & $0.002\pm0.64$\\
    40 & $4.23\pm73$ & $-0.010\pm0.31$ & $0.02\pm0.18$ & $0.06\pm3.10$ & $-0.023\pm0.33$ & $-0.012\pm0.16$ & $-0.008\pm0.18$ & $-0.002\pm0.66$\\
    35 & $3.95\pm81$ & $-0.005\pm0.34$ & $0.02\pm0.20$ & $-0.04\pm3.74$ & $-0.026\pm0.33$ & $-0.009\pm0.16$ & $-0.010\pm0.18$ & $-0.013\pm0.68$\\
    30 & $5.35\pm90$ & $-0.007\pm0.38$ & $0.02\pm0.20$ & $-0.02\pm3.88$ & $-0.030\pm0.35$ & $-0.015\pm0.17$ & $-0.013\pm0.19$ & $-0.008\pm0.71$\\
    26 & $7.29\pm100$ & $0.008\pm0.39$ & $0.03\pm0.23$ & $-0.06\pm4.41$ & $-0.031\pm0.36$ & $-0.016\pm0.18$ & $-0.015\pm0.20$ & $-0.020\pm0.74$\\
    23 & $7.88\pm111$ & $0.012\pm0.42$ & $0.04\pm0.25$ & $-0.07\pm5.01$ & $-0.035\pm0.37$ & $-0.017\pm0.19$ & $-0.014\pm0.21$ & $-0.031\pm0.76$\\
    20 & $11.4\pm127$ & $0.018\pm0.46$ & $0.05\pm0.28$ & $-0.13\pm5.93$ & $-0.036\pm0.38$ & $-0.017\pm0.19$ & $-0.014\pm0.21$ & $-0.035\pm0.79$\\
    18 & $12.4\pm138$ & $0.022\pm0.48$ & $0.06\pm0.30$ & $-0.10\pm6.42$ & $-0.036\pm0.39$ & $-0.018\pm0.20$ & $-0.014\pm0.22$ & $-0.057\pm0.80$\\
    16 & $17.7\pm149$ & $0.023\pm0.51$ & $0.07\pm0.33$ & $0.06\pm7.29$ & $-0.041\pm0.39$ & $-0.016\pm0.20$ & $-0.015\pm0.22$ & $-0.051\pm0.83$\\
    14 & $17.3\pm172$ & $0.014\pm0.55$ & $0.08\pm0.36$ & $-0.02\pm7.59$ & $-0.039\pm0.40$ & $-0.024\pm0.21$ & $-0.014\pm0.23$ & $-0.080\pm0.85$\\
    12 & $24.1\pm196$ & $0.013\pm0.60$ & $0.11\pm0.39$ & $0.09\pm8.28$ & $-0.034\pm0.42$ & $-0.022\pm0.21$ & $-0.013\pm0.23$ & $-0.113\pm0.89$\\
    10 & $17.5\pm233$ & $-0.002\pm0.64$ & $0.12\pm0.46$ & $0.23\pm9.59$ & $-0.017\pm0.43$ & $-0.026\pm0.22$ & $-0.014\pm0.24$ & $-0.163\pm0.94$\\
    \bottomrule
    \end{tabular}
\end{table*}

\begin{table*}[ht!]
    \centering
    \small
    \caption{Overview of the cINN predictive performance in terms of the NRMSE of the target parameters with respect to the ground truth on synthetic test data for different S/N. }
    \label{tab:nrmse_summary}
    \begin{tabular}{lcccccccc}
    \toprule
    \multicolumn{1}{c}{S/N} & $T_\mathrm{eff}$ & \multirow{2}{*}{$\log\left(\frac{g}{\mathrm{cms}^{-2}}\right)$} & \multirow{2}{*}{$[\mathrm{Fe}/\mathrm{H}]$} & $v_\mathrm{broad}$ & $v_\mathrm{mic}$ & \multicolumn{3}{c}{$[\mathrm{x}/\mathrm{Fe}]$} \\
    \multicolumn{1}{c}{(\AA$^{-1}$)} & $(K)$ & & & $(\mathrm{kms}^{-1})$ & $(\mathrm{kms}^{-1})$ & Ca & Mg & Li \\
    \midrule
    1000 & 0.008 & 0.028 & 0.02 & 0.012 & 0.18 & 0.071 & 0.066 & 0.054\\
    250 & 0.008 & 0.031 & 0.02 & 0.013 & 0.182 & 0.076 & 0.071 & 0.056\\
    180 & 0.009 & 0.033 & 0.021 & 0.013 & 0.184 & 0.079 & 0.075 & 0.057\\
    130 & 0.009 & 0.035 & 0.022 & 0.015 & 0.188 & 0.084 & 0.081 & 0.059\\
    100 & 0.011 & 0.039 & 0.023 & 0.016 & 0.191 & 0.09 & 0.087 & 0.06\\
    80 & 0.012 & 0.043 & 0.025 & 0.018 & 0.195 & 0.094 & 0.095 & 0.063\\
    50 & 0.015 & 0.054 & 0.029 & 0.026 & 0.207 & 0.111 & 0.11 & 0.069\\
    45 & 0.016 & 0.058 & 0.031 & 0.029 & 0.212 & 0.115 & 0.115 & 0.071\\
    40 & 0.018 & 0.059 & 0.031 & 0.031 & 0.217 & 0.12 & 0.119 & 0.073\\
    35 & 0.019 & 0.064 & 0.035 & 0.037 & 0.223 & 0.126 & 0.123 & 0.075\\
    30 & 0.022 & 0.07 & 0.036 & 0.039 & 0.231 & 0.132 & 0.129 & 0.078\\
    26 & 0.024 & 0.073 & 0.04 & 0.044 & 0.239 & 0.138 & 0.134 & 0.082\\
    23 & 0.027 & 0.079 & 0.044 & 0.05 & 0.244 & 0.144 & 0.138 & 0.083\\
    20 & 0.031 & 0.086 & 0.05 & 0.059 & 0.252 & 0.148 & 0.141 & 0.086\\
    18 & 0.033 & 0.091 & 0.053 & 0.064 & 0.258 & 0.151 & 0.146 & 0.089\\
    16 & 0.036 & 0.096 & 0.058 & 0.073 & 0.264 & 0.155 & 0.146 & 0.091\\
    14 & 0.041 & 0.102 & 0.064 & 0.076 & 0.27 & 0.162 & 0.151 & 0.094\\
    12 & 0.047 & 0.112 & 0.071 & 0.083 & 0.278 & 0.165 & 0.156 & 0.099\\
    10 & 0.056 & 0.12 & 0.083 & 0.096 & 0.287 & 0.17 & 0.161 & 0.105\\
    \bottomrule
    \end{tabular}
\end{table*}

\begin{figure*}
    \centering
    \includegraphics[width=\linewidth]{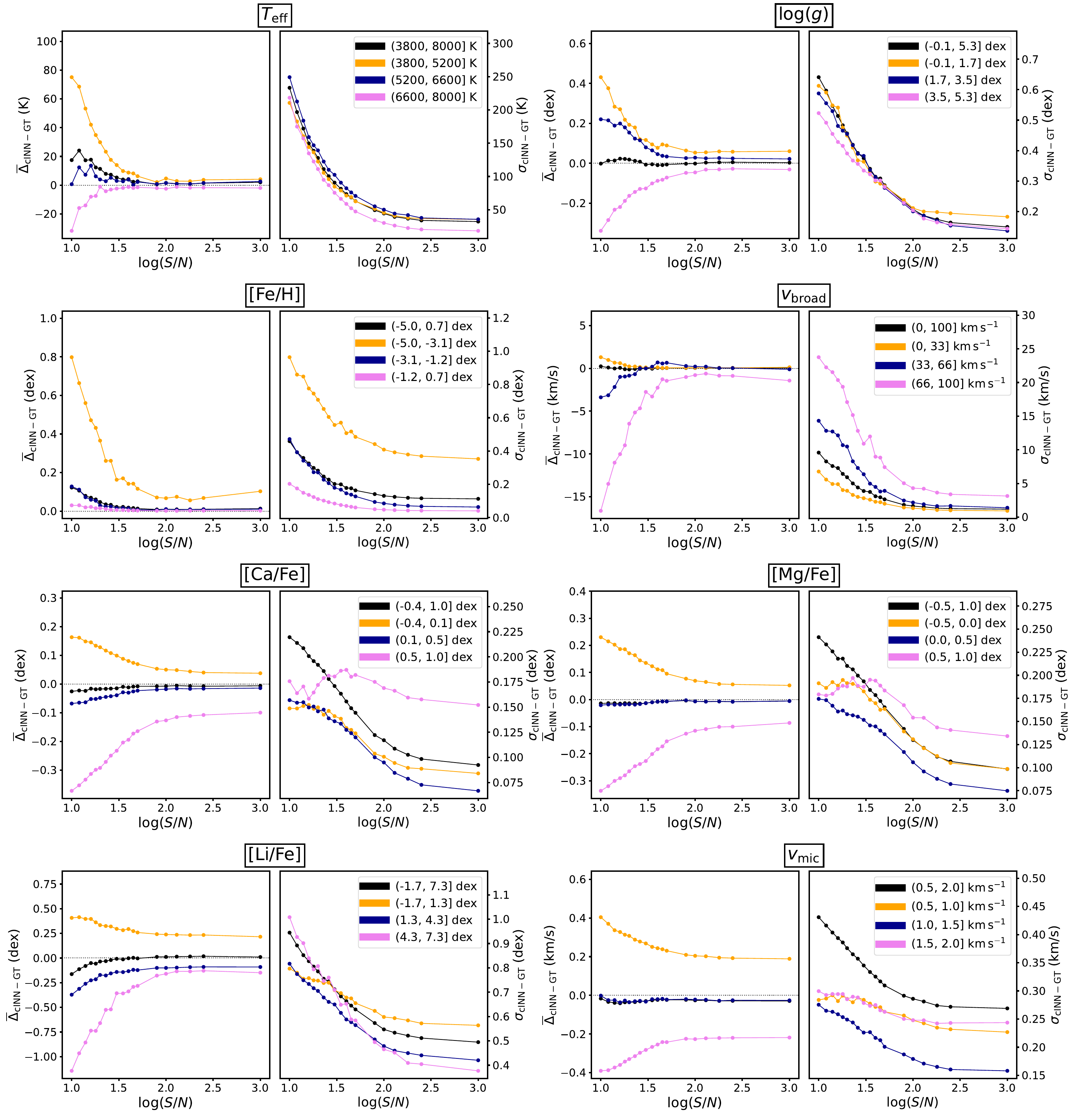}
    \caption{Summary of cINN performance of synthetic spectra as a function of S/N. Expanding upon Fig.~\ref{fig:Synth_ME_SIGMA_vs_SNR}, shown is a detailed breakdown of the mean residual $\overline{\Delta}_\mathrm{cINN-GT}$ and standard deviation $\sigma_\mathrm{cINN-GT}$ of each target parameter, distinguishing the lower third (orange), middle third (dark blue), and upper third (violet) of the respective range in the test data. For comparison, the black line shows the respective results over the full trained parameter range. The black dotted line indicates $\overline{\Delta}_\mathrm{cINN-GT} = 0$ for reference. }
    \label{fig:Synth_ME_SIGMA_vs_SNR_DETAILED}
\end{figure*}

\begin{figure*}
    \centering
    \includegraphics[width=\linewidth]{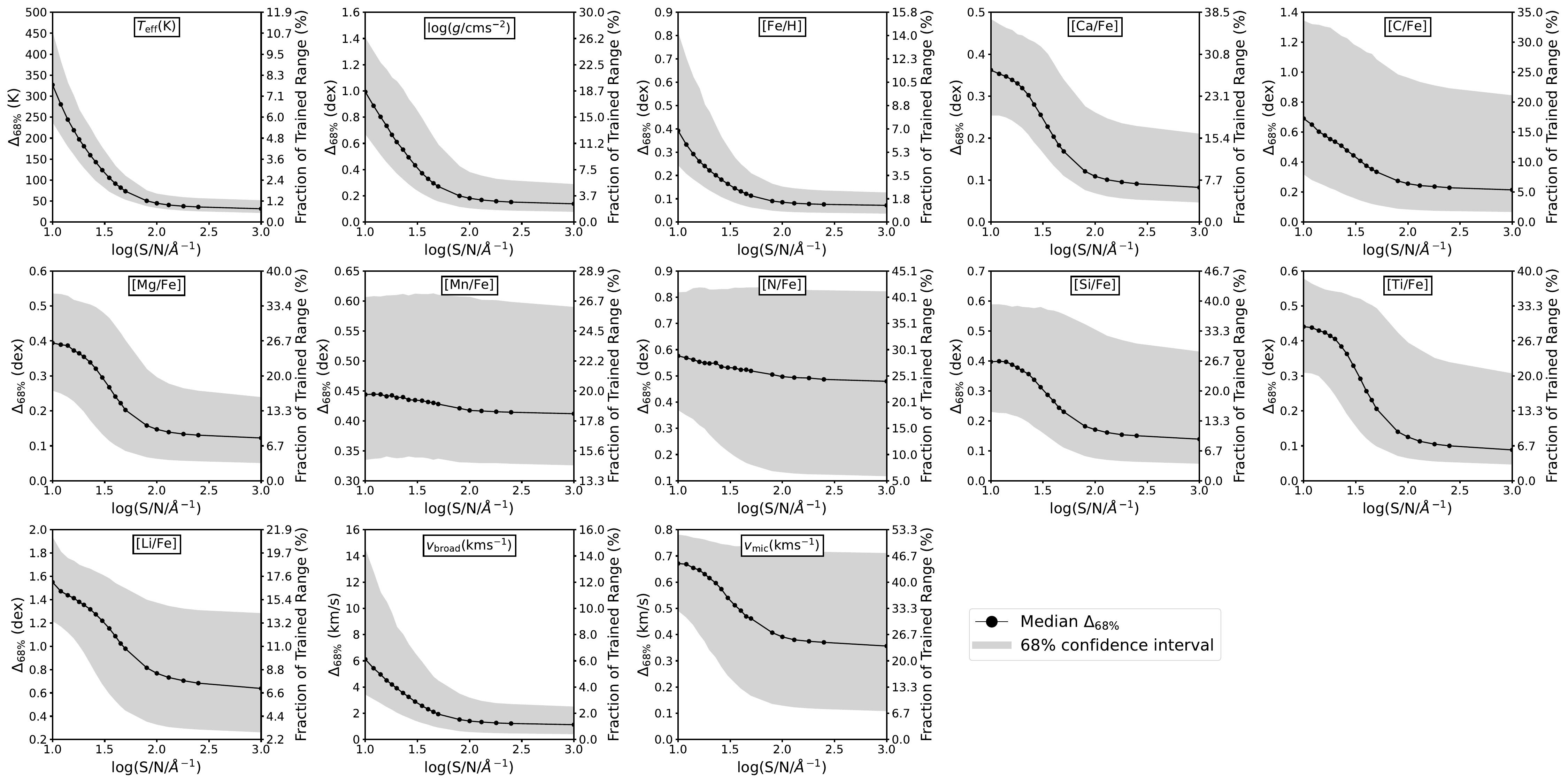}
    \caption{Overview of the median width of the 68\% confidence interval for the individual components of the predicted posterior distributions as a function of the S/N. The grey-shaded region indicates the 68\% confidence interval.}
    \label{fig:u68_vs_snr}
\end{figure*}

\begin{figure*}
    \centering
    \includegraphics[width=\linewidth]{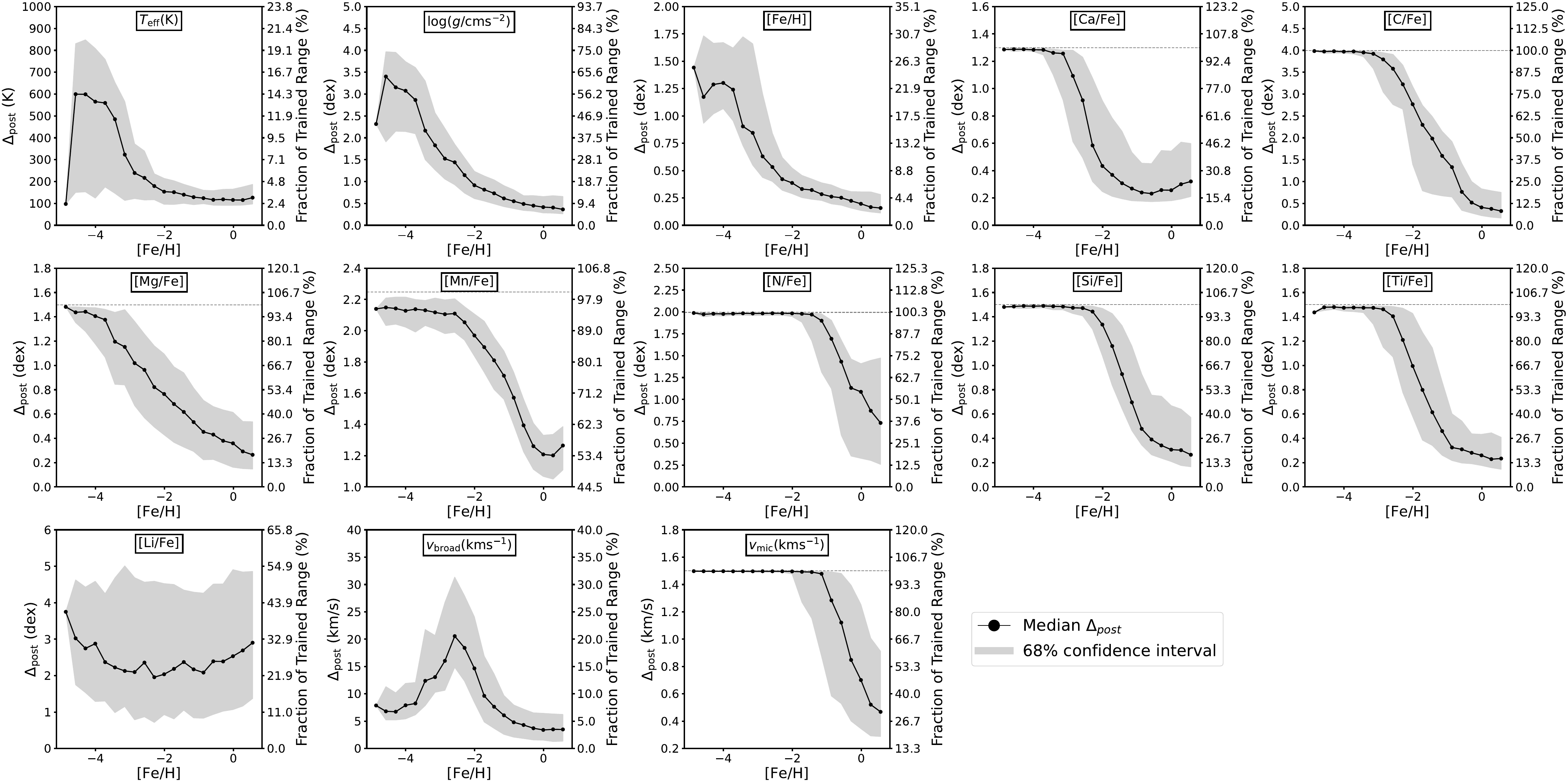}
    \caption{Overview of the median total width of the predicted posteriors for the individual components as a function of \feh~for $\mathrm{S/N} = 250\,$\AA$^{-1}$ on the synthetic test spectra. They grey shaded region indicates the 68\% confidence interval. The horizontal dashed line, indicates a full posterior width that corresponds to 100\% of the parameter range in the training set.}
    \label{fig:full_posterior_vs_feh}
\end{figure*}

\begin{figure*}
    \centering
    \includegraphics[width=0.87\linewidth]{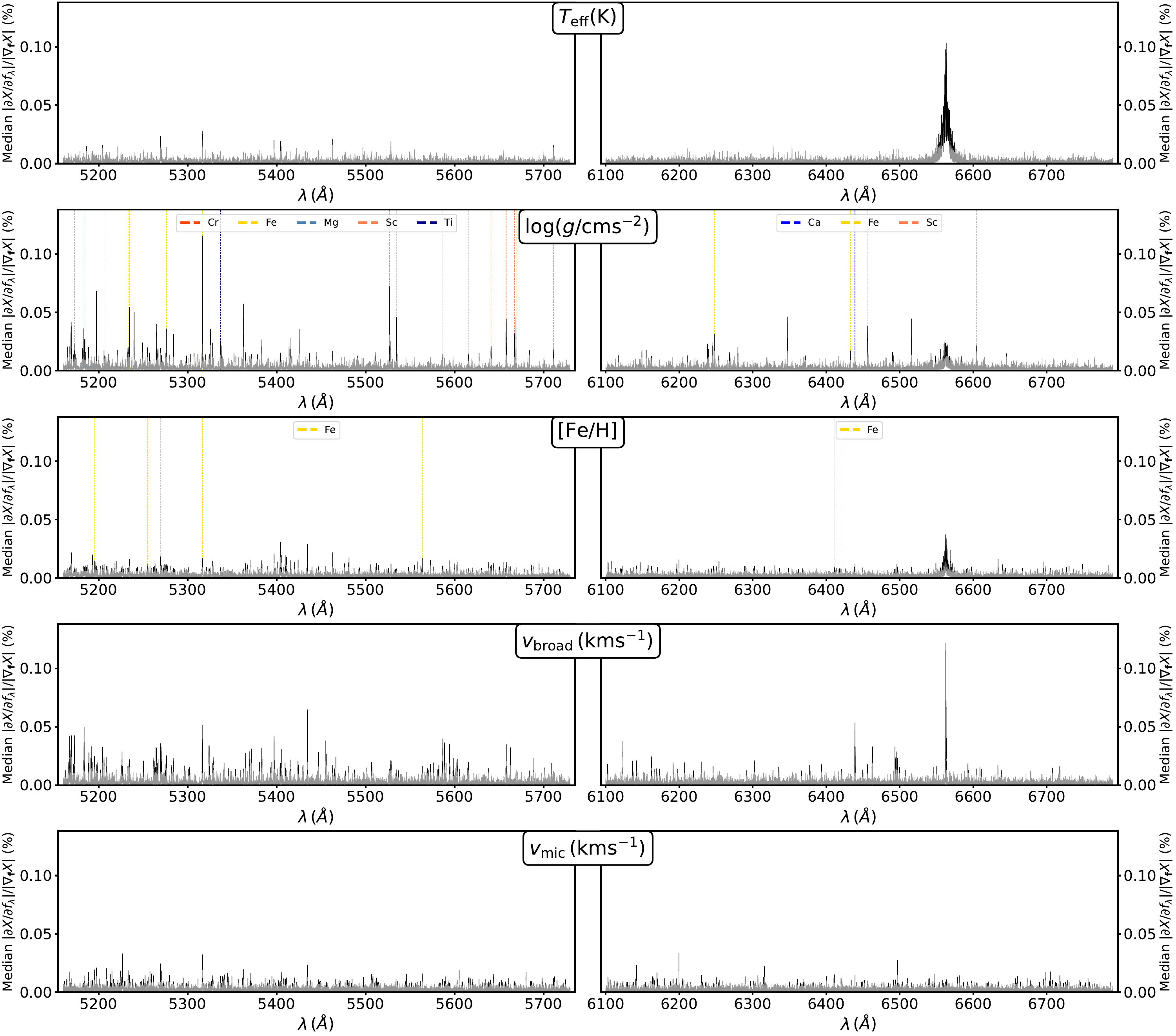}
    \caption{Average feature importance in the cINN for the primary stellar properties. Shown is the median relative contribution of each flux bin to the total gradient of each target parameter with respect to the total flux vector averaged over a synthetic test set with $\mathrm{S}/\mathrm{N} = 250\,$\AA$^{-1}$. Black lines indicate flux bins whose median contribution is larger than the mean by more than $3\sigma$. In each row, the left panel shows the green 4MOST window, while the right panel presents the red window. The dashed lines in the \logg~and \feh~panels indicate common diagnostic absorption lines of various elements that coincide with a peak in the feature importance.}
    \label{fig:saliency1}
\end{figure*}

\begin{figure*}
    \centering
    \includegraphics[width=0.87\linewidth]{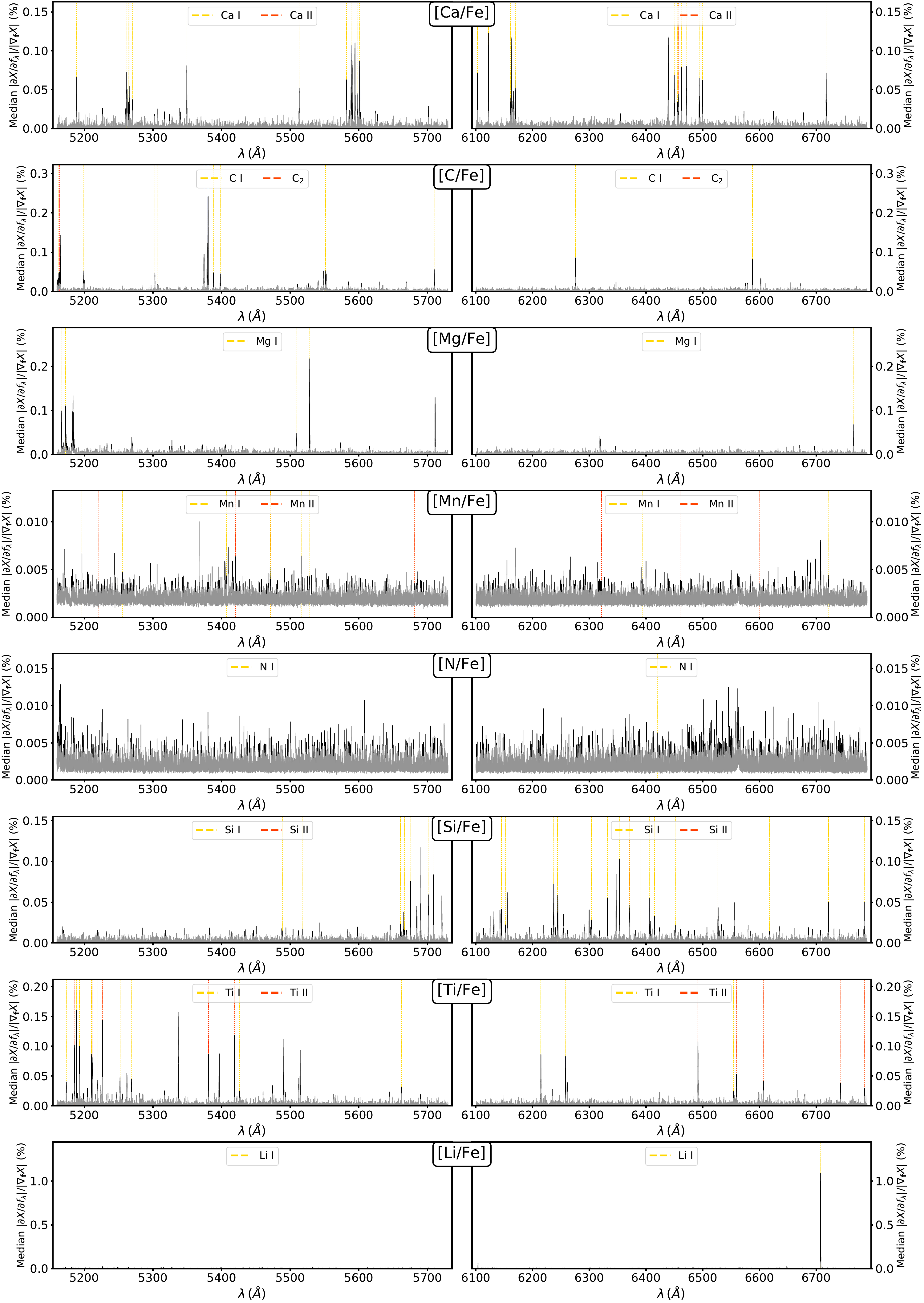}
    \caption{Average feature importance in the cINN for metal abundances. Shown is the median relative contribution of each flux bin to the total gradient of each target parameter with respect to the total flux vector averaged over a synthetic test set with $\mathrm{S}/\mathrm{N} = 250\,$\AA$^{-1}$. Black lines indicate flux bins whose median contribution is larger than the mean by more than $3\sigma$. In each row, the left panel shows the green 4MOST window, while the right panel presents the red window. The dashed lines indicate absorption lines of the respective elements that are included in the Turbospectrum models and coincide with peaks in the feature importance.}
    \label{fig:saliency2}
\end{figure*}

This appendix provides additional material related to the tests of the trained cINN on the held-out synthetic test spectra in Sect.~\ref{sec:results_synthetic}. 
Supplemental to Fig.~\ref{fig:Synth_ME_SIGMA_vs_SNR}, Tables~\ref{tab:me_summary} and \ref{tab:nrmse_summary} and provide summaries of the $\overline{\Delta}_\mathrm{cINN-GT}\pm\sigma_\mathrm{cINN-GT}$ and NRMSE performance metrics of the cINN MAP estimates on the synthetic test spectra across all S/N values that the cINN is trained for. 
Figure~\ref{fig:Synth_ME_SIGMA_vs_SNR_DETAILED} further expands upon Fig.~\ref{fig:Synth_ME_SIGMA_vs_SNR} by providing a breakdown of the mean residual $\overline{\Delta}_\mathrm{cINN-GT}$ and standard deviation $\sigma_\mathrm{cINN-GT}$ as a function of the S/N for different intervals of the respective target parameters. This figure shows that for most target parameters the trend of the mean residual is inverted between the lower third and upper third of the range. Specifically, there is a trend towards overestimation as the S/N decreases in the lower third, whereas in the upper third we observe a trend towards underestimation instead.  
Figure~\ref{fig:u68_vs_snr} shows the median width $\Delta_{68\%}$ of the 68\% confidence interval of the predicted posteriors for all of the target parameters of the cINN as a function of the S/N of the input spectrum. This figure demonstrates that the cINN learned to associate a larger uncertainty in the input fluxes with larger uncertainties in the predicted physical parameters. We also observe a notable increase in the posterior widths below $\mathrm{S/N} = 80\,$\AA$^{-1}$, rendering, in particular, the elemental abundances a lot harder to constrain. We note that a network specifically trained for the lower S/N regime might improve upon the performance of the general purpose model that is presented here.
Analogously to Fig.~\ref{fig:u68_vs_snr}, Fig.~\ref{fig:full_posterior_vs_feh} presents the median full width of the predicted posterior distributions as a function of \feh~for a fixed $\mathrm{S/N}=250\,$\AA$^{-1}$, highlighting a notable increase in the cINN intrinsic uncertainties in particular in the $[\mathrm{Fe/H}] < -2.5$ regime.
Figures~\ref{fig:saliency1} and \ref{fig:saliency2} shows a feature importance analysis on the trained network in the form of saliency maps for the target parameters. In particular, we computed the median network internal gradients $(\partial X/\partial f_\lambda)_i$ of each target parameter, $X$, with respect to the flux, $f_\lambda$, in each wavelength bin over all synthetic test spectra with $\mathrm{S/N} = 250\,$\AA$^{-1}$. To highlight the relative importance of each bin, the gradients $(\partial X/\partial f_\lambda)_i$ are normalised in Figs.~\ref{fig:saliency1} and \ref{fig:saliency2} to the magnitude of the total gradient $|\nabla_\mathrm{f}X|_i$ of $X$ with respect to the entire flux vector $\mathrm{f}$ for each test spectrum $i$ prior to taking the average. Bins with a median gradient that is larger by more than $3\sigma$ than the mean median gradient are highlighted in black. This analysis, in particular when looking at the elemental abundances, demonstrates that the cINN learned to pay attention to sections of the spectra that are also part of classical evaluation techniques for characterising spectra, as the highest peaks in saliency almost always coincide with absorption lines of the corresponding elements. 

\newpage
\section{Application to benchmark spectra}

\begin{table*}[h!]
    \centering
    \begin{threeparttable}
        \caption{Literature stellar properties of the set of sources used for validation.}
        \label{tab:benchmark_params}
        \scriptsize
        \begin{tabular}{lcrrrl}
        \toprule
        Star & SpType & $T_\mathrm{eff}$ (K) & $\log(g/\mathrm{cms}^{-2})$ & $[\mathrm{Fe}/\mathrm{H}]$ & Reference \\
        \midrule
        15PEG & F Dwarf (V)* & $5728 \pm 22$ & $4.28 \pm 0.03$ & $0.13 \pm 0.01$ &  {\cite{2021MNRAS.505.4496G}} \\
        18Sco & G Dwarf (V)$^\ddagger$ & $5824 \pm 30$ & $4.42 \pm 0.01$ & $0.06 \pm 0.01$ &  {\cite{2024A&A...682A.145S, Casamiquela2026}} \\
        31AQL & G Subgiant (IV)* & $5776 \pm 3$ & $4.28 \pm 0.01$ & $0.59 \pm 0.00$ &  {\cite{2021MNRAS.505.4496G}} \\
        61CygA & K Dwarf (V)$^\ddagger$ & $4398 \pm 34$ & $4.63 \pm 0.01$ & $-0.32 \pm 0.14$ &  {\cite{2024A&A...682A.145S, Casamiquela2026}} \\
        61CygB & K Dwarf (V)$^\ddagger$ & $4174 \pm 47$ & $4.68 \pm 0.02$ & $-0.59 \pm 0.21$ &  {\cite{2024A&A...682A.145S, Casamiquela2026}} \\
        Arcturus & K Giant (III)$^\ddagger$ & $4277 \pm 23$ & $1.58 \pm 0.07$ & $-0.55 \pm 0.04$ &  {\cite{2024A&A...682A.145S, Casamiquela2026}} \\
        BD+26\_3578 & B Subgiant (IV)$^\dagger$ & $6432 \pm 2$ & $3.75 \pm 0.01$ & $-2.02 \pm 0.02$ &  {\cite{2021MNRAS.505.4496G}} \\
        BD-04\_3208 & A Subgiant (IV)$^\dagger$ & $6223 \pm 50$ & $3.66 \pm 0.15$ & $-2.34 \pm 0.05$ &  {\cite{2022A&A...663A...4S}} \\
        G113-22 & G Subgiant (IV)$^\dagger$ & $5595 \pm 46$ & $3.70 \pm 0.37$ & $-1.07 \pm 0.09$ &  {\cite{2022A&A...663A...4S}} \\
        G165-39 & F Subdwarf (VI)* & $6333 \pm 2$ & $3.82 \pm 0.01$ & $-1.90 \pm 0.02$ &  {\cite{2021MNRAS.505.4496G}} \\
        G207-5 & F Dwarf (V)$^\dagger$ & $6003 \pm 2$ & $4.18 \pm 0.01$ & $-0.30 \pm 0.01$ &  {\cite{2021MNRAS.505.4496G}} \\
        G28-43 & G Subdwarf (VI)* & $4923 \pm 34$ & $4.58 \pm 0.07$ & $-1.56 \pm 0.02$ &  {\cite{2022ApJS..259...35A}} \\
        G41-41 & A Subdwarf (VI)* & $6455 \pm 6$ & $4.07 \pm 0.03$ & $-3.09 \pm 0.66$ &  {\cite{2021MNRAS.505.4496G}} \\
        G48-29 & A Subdwarf (VI)* & $6279 \pm 111$ & $3.99 \pm 0.27$ & $-2.52 \pm 0.24$ &  {\cite{2022A&A...663A...4S}} \\
        G63-46 & F Dwarf (V)* & $5719 \pm 42$ & $3.89 \pm 0.17$ & $-0.89 \pm 0.07$ &  {\cite{2022A&A...663A...4S}} \\
        G64-12 & F Subdwarf (VI)* & $6368 \pm 40$ & $4.11 \pm 0.07$ & $-3.26 \pm 0.05$ &  {\cite{2022A&A...663A...4S}} \\
        G78-1 & F Dwarf (V)$^\dagger$ & $5995 \pm 3$ & $4.11 \pm 0.02$ & $-0.66 \pm 0.02$ &  {\cite{2021MNRAS.505.4496G}} \\
        G9-16 & F Dwarf (V)* & $6858 \pm -$ & $4.28 \pm -$ & $-1.20 \pm -$ &  {\cite{2023ApJS..266...41P}} \\
        Gmb1830 & G Dwarf (V)$^\ddagger$ & $5235 \pm 18$ & $4.72 \pm 0.01$ & $-1.20 \pm 0.06$ &  {\cite{2024A&A...682A.145S, Casamiquela2026}} \\
        HD101177 & G Dwarf (V)* & $5887 \pm 70$ & $4.43 \pm 0.07$ & $-0.26 \pm 0.05$ &  {\cite{2022A&A...663A...4S}} \\
        HD102158 & G Dwarf (V)$^\dagger$ & $5761 \pm 20$ & $4.32 \pm 0.04$ & $-0.43 \pm 0.02$ &  {\cite{2022A&A...663A...4S}} \\
        HD103095 & K Dwarf (V)* & $4991 \pm 30$ & $4.60 \pm 0.05$ & $-1.26 \pm 0.02$ &  {\cite{2022ApJS..259...35A}} \\
        HD104304 & G Subgiant (IV)* & $5441 \pm 6$ & $4.18 \pm 0.01$ & $0.32 \pm 0.01$ &  {\cite{2021MNRAS.505.4496G}} \\
        HD105631 & G Dwarf (V)* & $5402 \pm 19$ & $4.49 \pm 0.04$ & $0.18 \pm 0.02$ &  {\cite{2022A&A...663A...4S}} \\
        HD106210 & G Dwarf (V)* & $5664 \pm 20$ & $4.41 \pm 0.05$ & $-0.10 \pm 0.04$ &  {\cite{2022A&A...663A...4S}} \\
        HD106516 & F Dwarf (V)* & $6274 \pm 4$ & $4.47 \pm 0.03$ & $-0.82 \pm 0.02$ &  {\cite{2021MNRAS.505.4496G}} \\
        HD10697 & G Dwarf (V)* & $5610 \pm 39$ & $3.87 \pm 0.03$ & $0.16 \pm 0.01$ &  {\cite{2022ApJS..259...35A}} \\
        HD10700 & G Dwarf (V)* & $5374 \pm 2$ & $4.35 \pm 0.01$ & $-0.19 \pm 0.01$ &  {\cite{2021MNRAS.505.4496G}} \\
        HD107328 & K Giant (III)$^\ddagger$ & $4576 \pm 87$ & $1.77 \pm 0.22$ & $-0.36 \pm 0.06$ &  {\cite{2024A&A...682A.145S, Casamiquela2026}} \\
        HD110897 & F Dwarf (V)* & $5872 \pm 3$ & $4.25 \pm 0.02$ & $-0.62 \pm 0.01$ &  {\cite{2021MNRAS.505.4496G}} \\
        HD111395 & G Dwarf (V)* & $6279 \pm 15$ & $3.56 \pm 0.03$ & $-0.09 \pm 0.02$ &  {\cite{2020AJ....160...83S}} \\
        HD113139 & F Dwarf (V)* & $6800 \pm -$ & $4.10 \pm -$ & $-0.20 \pm -$ &  {\cite{2016A&A...589A..83G}} \\
        HD114174 & G Subgiant (IV)* & $5854 \pm -$ & $4.53 \pm -$ & $-0.02 \pm -$ &  {\cite{2021MNRAS.505.4496G}} \\
        HD122563 & K Bright Giant (II)$^\ddagger$ & $4642 \pm 35$ & $1.32 \pm 0.03$ & $-2.66 \pm 0.05$ &  {\cite{2024A&A...682A.145S, Casamiquela2026}} \\
        HD140283 & G Subgiant (IV)$^\ddagger$ & $5788 \pm 45$ & $3.75 \pm 0.09$ & $-2.41 \pm 0.04$ &  {\cite{2024A&A...682A.145S, Casamiquela2026}} \\
        HD144284 & F Dwarf (V)* & $6208 \pm 48$ & $3.79 \pm -$ & $0.19 \pm -$ &  {\cite{2017AJ....153...21L}} \\
        HD162003 & F Subgiant/Dwarf (IV/V)* & $6567 \pm 86$ & $4.06 \pm 0.32$ & $-0.02 \pm 0.11$ &  {\cite{2020AJ....159...90S}} \\
        HD164259 & F Dwarf (V)* & $6815 \pm 63$ & $4.11 \pm 0.02$ & $-0.10 \pm 0.05$ &  {\cite{2018A&A...614A..55A}} \\
        HD220009 & K Giant (III)$^\ddagger$ & $4227 \pm 77$ & $1.66 \pm 0.04$ & $-0.71 \pm 0.04$ &  {\cite{2024A&A...682A.145S, Casamiquela2026}} \\
        HD22879 & G Dwarf (V)$^\ddagger$ & $5962 \pm 86$ & $4.28 \pm 0.04$ & $-0.78 \pm 0.03$ &  {\cite{2024A&A...682A.145S, Casamiquela2026}} \\
        HD284248 & G Subdwarf (VI)* & $6253 \pm 3$ & $4.04 \pm 0.02$ & $-1.52 \pm 0.03$ &  {\cite{2021MNRAS.505.4496G}} \\
        HD84937 & F Dwarf (V)$^\ddagger$ & $6484 \pm 106$ & $4.16 \pm 0.05$ & $-1.97 \pm 0.07$ &  {\cite{2024A&A...682A.145S, Casamiquela2026}} \\
        OMI\_AQL & F Dwarf (V)* & $6249 \pm 3$ & $4.04 \pm 0.01$ & $0.28 \pm 0.01$ &  {\cite{2021MNRAS.505.4496G}} \\
        Procyon & F Subgiant (IV)$^\ddagger$ & $6582 \pm 5$ & $3.98 \pm 0.02$ & $-0.03 \pm 0.02$ &  {\cite{2024A&A...682A.145S, Casamiquela2026}} \\
        Sun & G Dwarf (V)$^\ddagger$ & $5771 \pm 10$ & $4.44 \pm 0.01$ & $0.00 \pm 0.00$ &  {\cite{2024A&A...682A.145S, Casamiquela2026}} \\
        $\alpha$ CenA & G Dwarf (V)$^\ddagger$ & $5804 \pm 13$ & $4.29 \pm 0.01$ & $0.22 \pm 0.03$ &  {\cite{2024A&A...682A.145S, Casamiquela2026}} \\
        $\alpha$ Cet & K Bright Giant (II)$^\ddagger$ & $3738 \pm 170$ & $0.66 \pm 0.07$ & $-0.57 \pm 0.17$ &  {\cite{2024A&A...682A.145S, Casamiquela2026}} \\
        $\alpha$ Tau & K Bright Giant (II)$^\ddagger$ & $3921 \pm 80$ & $1.17 \pm 0.05$ & $-0.24 \pm 0.11$ &  {\cite{2024A&A...682A.145S, Casamiquela2026}} \\
        $\beta$ Hyi & G Subgiant (IV)$^\ddagger$ & $5917 \pm 25$ & $3.97 \pm 0.04$ & $-0.06 \pm 0.03$ &  {\cite{2024A&A...682A.145S, Casamiquela2026}} \\
        $\beta$ Vir & F Dwarf (V)$^\ddagger$ & $6093 \pm 13$ & $4.08 \pm 0.02$ & $0.13 \pm 0.02$ &  {\cite{2024A&A...682A.145S, Casamiquela2026}} \\
        $\delta$ Eri & K Subgiant (IV)$^\ddagger$ & $4954 \pm 38$ & $3.79 \pm 0.10$ & $0.04 \pm 0.04$ &  {\cite{2024A&A...682A.145S, Casamiquela2026}} \\
        $\epsilon$ Eri & K Dwarf (V)$^\ddagger$ & $5062 \pm 29$ & $4.60 \pm 0.01$ & $-0.10 \pm 0.03$ &  {\cite{2024A&A...682A.145S, Casamiquela2026}} \\
        $\epsilon$ Vir & K Giant (III)$^\ddagger$ & $4950 \pm 10$ & $2.72 \pm 0.02$ & $-0.03 \pm 0.03$ &  {\cite{2024A&A...682A.145S, Casamiquela2026}} \\
        $\eta$ Boo & F Subgiant (IV)$^\ddagger$ & $6161 \pm 18$ & $3.82 \pm 0.05$ & $0.32 \pm 0.04$ &  {\cite{2024A&A...682A.145S, Casamiquela2026}} \\
        $\mu$ Ara & G Dwarf (V)$^\ddagger$ & $5974 \pm 61$ & $4.30 \pm 0.03$ & $0.41 \pm 0.04$ &  {\cite{2024A&A...682A.145S, Casamiquela2026}} \\
        $\mu$ Cas & G Dwarf (V)$^\ddagger$ & $5358 \pm 31$ & $4.49 \pm 0.01$ & $-0.81 \pm 0.04$ &  {\cite{2024A&A...682A.145S, Casamiquela2026}} \\
        $\mu$ Leo & K Giant (III)$^\ddagger$ & $4519 \pm 23$ & $2.43 \pm 0.06$ & $0.19 \pm 0.06$ &  {\cite{2024A&A...682A.145S, Casamiquela2026}} \\
        $\tau$ Cet & G Dwarf (V)$^\ddagger$ & $5397 \pm 15$ & $4.50 \pm 0.01$ & $-0.47 \pm 0.03$ &  {\cite{2024A&A...682A.145S, Casamiquela2026}} \\
        \bottomrule
        \end{tabular}
            \begin{tablenotes}
            \item [$*$] From \url{https://simbad.cds.unistra.fr/simbad/}. 
            \item [$^\dagger$] Spectral type from \url{https://simbad.cds.unistra.fr/simbad/}, luminosity class estimated based on \logg.
            \item [$^\ddagger$] From \url{https://www.blancocuaresma.com/s/benchmarkstars}, i.e.~\cite{2024A&A...682A.145S, Casamiquela2026}.
        \end{tablenotes}

    \end{threeparttable}
\end{table*}

\begin{table*}
    \centering
    \small
    \caption{List of absorption lines used to fit abundances with TSFitPy.}
    \label{tab:tsfitpy_lines}
    \begin{tabular}{lccccclccccc}
    \toprule
    Element & $\lambda$ [\AA] & $E_{low}$ [eV] & $\log(gf)$ & $g_{up}$ & Isotope & Element & $\lambda$ [\AA] & $E_{low}$ [eV] & $\log(gf)$ & $g_{up}$ & Isotope \\
    \midrule
    Ca I & 5188.844 & 2.933 & -0.075 & 5 &  & Fe I & 5560.211 & 4.435 & -1.090 & 7 &  \\
    Ca I & 5349.465 & 2.709 & -0.310 & 7 &  & Fe I & 5638.262 & 4.220 & -0.720 & 7 &  \\
    Ca I & 5581.965 & 2.523 & -0.555 & 7 &  & Fe I & 5661.345 & 4.284 & -1.756 & 3 &  \\
    Ca I & 5588.749 & 2.526 & 0.358 & 7 &  & Fe I & 5679.023 & 4.652 & -0.820 & 7 &  \\
    Ca I & 5590.114 & 2.521 & -0.571 & 5 &  & Fe I & 6165.360 & 4.143 & -1.473 & 9 &  \\
    Ca I & 5594.462 & 2.523 & 0.097 & 5 &  & Fe I & 6187.989 & 3.943 & -1.620 & 9 &  \\
    Ca I & 5601.277 & 2.526 & -0.523 & 5 &  & Fe I & 6270.224 & 2.858 & -2.470 & 3 &  \\
    Ca I & 6102.723 & 1.879 & -0.850 & 3 &  & Fe II & 5234.623 & 3.221 & -2.180 & 6 &  \\
    Ca I & 6122.217 & 1.886 & -0.380 & 3 &  & Fe II & 5325.552 & 3.221 & -3.160 & 8 &  \\
    Ca I & 6162.173 & 1.899 & -0.170 & 3 &  & Fe II & 5425.249 & 3.199 & -3.220 & 10 &  \\
    Ca I & 6169.563 & 2.526 & -0.478 & 5 &  & Fe II & 6456.379 & 3.903 & -2.185 & 6 &  \\
    Ca I & 6439.075 & 2.526 & 0.390 & 9 &  & Li I & 6707.764 & 0.000 & -0.002 & 4 & 7 \\
    Ca I & 6449.808 & 2.521 & -0.502 & 5 &  & Li I & 6707.915 & 0.000 & -0.303 & 2 & 7 \\
    Ca I & 6462.567 & 2.523 & 0.262 & 7 &  & Li I & 6707.921 & 0.000 & -0.002 & 4 & 6 \\
    Ca I & 6493.781 & 2.521 & -0.109 & 5 &  & Li I & 6708.072 & 0.000 & -0.303 & 2 & 6 \\
    Ca I & 6717.681 & 2.709 & -0.524 & 3 &  & Mg I & 5167.322 & 2.709 & -0.931 & 3 &  \\
    Fe I & 5242.491 & 3.634 & -0.967 & 11 &  & Mg I & 5172.684 & 2.712 & -0.363 & 3 &  \\
    Fe I & 5365.399 & 3.573 & -1.020 & 9 &  & Mg I & 5183.604 & 2.717 & -0.168 & 3 &  \\
    Fe I & 5379.574 & 3.695 & -1.514 & 11 &  & Mg I & 5528.405 & 4.346 & -0.547 & 5 &  \\
    Fe I & 5398.279 & 4.446 & -0.630 & 5 &  &  Mg I & 5711.088 & 4.346 & -1.742 & 1 &  \\
    Fe I & 5543.936 & 4.218 & -1.040 & 5 &  & & & & & & \\
    \bottomrule
    \end{tabular}
\end{table*}

This appendix contains supplementary material regarding the application of the trained cINN to real spectra in Sect. \ref{sec:results_benchmark}. 
Table~\ref{tab:benchmark_params} lists the set of sources, for which we analysed spectra in this application of our cINN, along with literature results of \teff, \logg,~and \feh~(with the corresponding references).
Table~\ref{tab:tsfitpy_lines} summarises the set of absorption lines used in Sect.~\ref{sec:results_benchmark} to determine the [Fe/H], [Ca/Fe], [Mg/Fe] and[Li/Fe] abundance ratios with TSFitPy for our set of real benchmark spectra.  
Table~\ref{tab:benchmark_predictions} summarises the TSFitPy abundances determined for our set of real benchmark spectra as outlined in Section~\ref{sec:results_benchmark}. It also lists the final cINN parameter estimates for all benchmark spectra after the bias correction that is detailed in the following Sect.~\ref{app:Bias_corr}.
Table~\ref{tab:benchmark_predictions_raw} provides an overview of the raw MAP estimates of the cINN for all 13 target parameters and the sample rejection fraction $f_\mathrm{reject}$ across the entire set of real spectra analysed in Sect.~\ref{sec:results_benchmark}.

\begin{table*}[h!]
    \centering
    \scriptsize
    \begin{threeparttable}[b]
        \caption{Overview of cINN MAP predictions and TSFitPy abundances for the benchmark spectra.}
        \label{tab:benchmark_predictions}
        \begin{tabular}{lccccccccccc}
        \toprule
        & \multicolumn{6}{c}{Final cINN prediction} & & \multicolumn{4}{c}{Turbospectrum fit$^*$} \\
        \cmidrule(rl){2-7}\cmidrule(rl){9-12}
        Star (Spectrum) &$T_\mathrm{eff}$ (K) & $\log\left(\frac{g}{\mathrm{cms}^{-2}}\right)$ & $[\mathrm{Fe/H}]$ & $[\mathrm{Ca/Fe}]$ & $[\mathrm{Mg/Fe}]$ & $[\mathrm{Li/Fe}]$ & $f_\mathrm{reject}$ & $[\mathrm{Fe/H}]$ & $[\mathrm{Ca/Fe}]$ & $[\mathrm{Mg/Fe}]$ & $[\mathrm{Li/Fe}]$\\
        \midrule
        15PEG (FOCES) & $6539$ & $3.78$ & $-0.58$ & $0.13$ & $0.3$ & $1.61$ & $0.13$ & $-0.57 \pm 0.16$ & $0.08 \pm 0.13$ & $0.44 \pm 0.05$ & - \\
        18Sco (NARVAL) & $5749$ & $4.36$ & $-0.1$ & $0.12$ & $0.09$ & $0.68$ & $0$ & $-0.09 \pm 0.07$ & $0.10 \pm 0.04$ & $0.06 \pm 0.01$ & $0.62 \pm 0.00$ \\
        31AQL (FOCES) & $5523$ & $4.2$ & $0.23$ & $0.06$ & $0.12$ & $-0.01$ & $0$ & $0.19 \pm 0.07$ & $-0.07 \pm 0.09$ & $0.03 \pm 0.03$ & - \\
        61CygA (NARVAL) & $4466$ & $4.54$ & $-0.39$ & $0.15$ & $0.27$ & $-0.2$ & $0.5$ & $-0.45 \pm 0.13$ & $0.13 \pm 0.06$ & $0.12 \pm 0.15$ & - \\
        61CygB (NARVAL) & $4208$ & $4.39$ & $-0.6$ & $0.39$ & $0.13$ & $-0.51$ & $0.72$ & $-0.68 \pm 0.01$ & $0.07 \pm 0.08$ & - & - \\
        Arcturus (UVES) & $4369$ & $1.7$ & $-0.62$ & $0.09$ & $0.44$ & $-0.39$ & $0.01$ & $-0.68 \pm 0.07$ & $0.19 \pm 0.08$ & $0.50 \pm 0.00$ & - \\
        Arcturus (NARVAL) & $4347$ & $1.63$ & $-0.65$ & $0.14$ & $0.43$ & $-0.29$ & $0$ & $-0.73 \pm 0.06$ & $0.19 \pm 0.11$ & $0.48 \pm 0.00$ & - \\
        Arcturus (UVES) & $4381$ & $1.76$ & $-0.61$ & $0.05$ & $0.37$ & $-0.17$ & $0.01$ & $-0.63 \pm 0.07$ & $0.12 \pm 0.15$ & $0.45 \pm 0.03$ & - \\
        BD+26\_3578 (FOCES) & $6500$ & $4.32$ & $-2.23$ & $0.4$ & $0.33$ & $3.69$ & $0.22$ & $-2.26 \pm 0.00$ & $0.40 \pm 0.07$ & $0.35 \pm 0.01$ & $3.55 \pm 0.00$ \\
        BD-04\_3208 (FOCES) & $6385$ & $4.32$ & $-2.32$ & $0.33$ & $0.38$ & $3.57$ & $0.16$ & $-2.21 \pm 0.08$ & $0.40 \pm 0.07$ & $0.37 \pm 0.03$ & $3.53 \pm 0.00$ \\
        G113-22 (FOCES) & $5692$ & $3.77$ & $-1.14$ & $0.27$ & $0.47$ & $2.03$ & $0.1$ & $-1.11 \pm 0.11$ & $0.26 \pm 0.11$ & $0.44 \pm 0.04$ & $2.01 \pm 0.00$ \\
        G165-39 (FOCES) & $6418$ & $4.17$ & $-2$ & $0.37$ & $0.28$ & $3.18$ & $0.17$ & $-2.01 \pm 0.07$ & $0.37 \pm 0.06$ & $0.30 \pm 0.02$ & $3.23 \pm 0.00$ \\
        G207-5 (FOCES) & $5904$ & $4.28$ & $-0.5$ & $0.06$ & $0.18$ & $1.49$ & $0.36$ & $-0.46 \pm 0.08$ & $0.10 \pm 0.08$ & $0.16 \pm 0.02$ & $1.67 \pm 0.00$ \\
        G28-43 (FOCES) & $5096$ & $4.64$ & $-1.7$ & $0.32$ & $0.37$ & $1.9$ & $0.35$ & $-1.58 \pm 0.05$ & $0.33 \pm 0.08$ & $0.28 \pm 0.08$ & $1.81 \pm 0.00$ \\
        G41-41 (FOCES) & $6517$ & $4.21$ & $-2.73$ & $0.19$ & $0.31$ & $4.07$ & $0.22$ & $-2.71 \pm 0.00$ & $0.39 \pm 0.10$ & $0.22 \pm 0.05$ & $3.91 \pm 0.00$ \\
        G48-29 (FOCES) & $6517$ & $4.72$ & $-2.83$ & $0.29$ & $0.37$ & $3.96$ & $0.3$ & $-2.50 \pm 0.00$ & $0.49 \pm 0.02$ & $0.39 \pm 0.07$ & $3.98 \pm 0.00$ \\
        G63-46 (FOCES) & $5801$ & $4.19$ & $-0.79$ & $0.25$ & $0.43$ & $1.04$ & $0.23$ & $-0.77 \pm 0.07$ & $0.28 \pm 0.10$ & $0.40 \pm 0.03$ & $1.09 \pm 0.00$ \\
        G64-12 (FOCES) & $6403$ & $4.93$ & $-3.71$ & $0.29$ & $0.72$ & $5.14$ & $0.3$ & $-3.20 \pm 0.00$ & $0.98 \pm 0.13$ & $0.68 \pm 0.01$ & - \\
        G78-1 (FOCES) & $5978$ & $4.12$ & $-0.92$ & $0.25$ & $0.46$ & $1.98$ & $0.29$ & $-0.89 \pm 0.10$ & $0.28 \pm 0.13$ & $0.47 \pm 0.02$ & $1.81 \pm 0.00$ \\
        G9-16 (FOCES) & $6922$ & $4.47$ & $-1.41$ & $0.4$ & $0.41$ & $2.96$ & $0.13$ & $-1.37 \pm 0.05$ & $0.37 \pm 0.07$ & $0.45 \pm 0.04$ & - \\
        Gmb1830 (NARVAL) & $5037$ & $4.58$ & $-1.41$ & $0.35$ & $0.42$ & $1.4$ & $0.18$ & $-1.37 \pm 0.08$ & $0.32 \pm 0.03$ & $0.24 \pm 0.06$ & - \\
        HD101177 (FOCES) & $5897$ & $4.27$ & $-0.31$ & $0.08$ & $0.18$ & $1.46$ & $0.02$ & $-0.28 \pm 0.08$ & $0.09 \pm 0.08$ & $0.11 \pm 0.02$ & $1.43 \pm 0.00$ \\
        HD102158 (FOCES) & $5701$ & $4.02$ & $-0.5$ & $0.28$ & $0.42$ & $1.15$ & $0.28$ & $-0.50 \pm 0.06$ & $0.27 \pm 0.10$ & $0.43 \pm 0.02$ & $0.54 \pm 0.00$ \\
        HD103095 (FOCES) & $5076$ & $4.64$ & $-1.41$ & $0.33$ & $0.41$ & $1.37$ & $0.23$ & $-1.34 \pm 0.06$ & $0.32 \pm 0.05$ & $0.32 \pm 0.08$ & - \\
        HD104304 (FOCES) & $5467$ & $4.48$ & $0.07$ & $0.09$ & $0.12$ & $0.11$ & $0$ & $0.06 \pm 0.06$ & $-0.02 \pm 0.10$ & $0.04 \pm 0.02$ & - \\
        HD105631 (FOCES) & $5346$ & $4.55$ & $-0.05$ & $0.1$ & $0.09$ & $0.38$ & $0$ & $-0.04 \pm 0.06$ & $0.08 \pm 0.09$ & $0.06 \pm 0.03$ & - \\
        HD106210 (FOCES) & $5627$ & $4.28$ & $-0.26$ & $0.15$ & $0.3$ & $0.96$ & $0$ & $-0.26 \pm 0.07$ & $0.17 \pm 0.06$ & $0.27 \pm 0.02$ & - \\
        HD106516 (FOCES) & $6368$ & $4.27$ & $-0.75$ & $0.23$ & $0.56$ & $1.99$ & $0.24$ & $-0.68 \pm 0.09$ & $0.24 \pm 0.10$ & $0.54 \pm 0.03$ & - \\
        HD10697 (FOCES) & $5555$ & $3.86$ & $-0.03$ & $0.06$ & $0.06$ & $0.95$ & $0$ & $0.02 \pm 0.07$ & $0.08 \pm 0.07$ & $0.09 \pm 0.01$ & $0.84 \pm 0.00$ \\
        HD10700 (FOCES) & $5298$ & $4.45$ & $-0.64$ & $0.24$ & $0.41$ & $0.86$ & $0.05$ & $-0.62 \pm 0.06$ & $0.26 \pm 0.06$ & $0.34 \pm 0.05$ & - \\
        HD107328 (NARVAL) & $4485$ & $1.95$ & $-0.55$ & $0.08$ & $0.38$ & $-0.27$ & $0$ & $-0.57 \pm 0.06$ & $0.10 \pm 0.10$ & $0.50 \pm 0.04$ & - \\
        HD107328 (UVES) & $4508$ & $2.05$ & $-0.53$ & $0.06$ & $0.29$ & $-0.14$ & $0$ & $-0.56 \pm 0.05$ & $0.08 \pm 0.08$ & $0.50 \pm 0.12$ & - \\
        HD110897 (FOCES) & $5855$ & $4.18$ & $-0.64$ & $0.11$ & $0.27$ & $1.21$ & $0.04$ & $-0.62 \pm 0.09$ & $0.11 \pm 0.09$ & $0.25 \pm 0.03$ & $1.52 \pm 0.00$ \\
        HD111395 (FOCES) & $5525$ & $4.38$ & $-0.08$ & $0.12$ & $0.07$ & $0.46$ & $0$ & $-0.02 \pm 0.05$ & $0.12 \pm 0.08$ & $0.07 \pm 0.01$ & - \\
        HD113139 (FOCES) & $6980$ & $4.28$ & $0.06$ & $0.14$ & $0.03$ & $1.71$ & $0.08$ & $-0.15 \pm 0.00$ & - & $0.05 \pm 0.00$ & - \\
        HD114174 (FOCES) & $5676$ & $4.27$ & $-0.1$ & $0.11$ & $0.16$ & $0.29$ & $0$ & $-0.10 \pm 0.06$ & $0.09 \pm 0.06$ & $0.13 \pm 0.01$ & - \\
        HD122563 (NARVAL) & $4847$ & $1.47$ & $-2.32$ & $0.09$ & $0.14$ & $1.91$ & $0.22$ & $-2.39 \pm 0.06$ & $0.11 \pm 0.08$ & $0.24 \pm 0.08$ & - \\
        HD140283 (UVES) & $6000$ & $4.03$ & $-2.31$ & $0.1$ & $0.26$ & $3.62$ & $0.21$ & $-2.33 \pm 0.06$ & $0.10 \pm 0.05$ & $0.19 \pm 0.03$ & $3.62 \pm 0.00$ \\
        HD140283 (NARVAL) & $5876$ & $3.74$ & $-2.38$ & $0.11$ & $0.25$ & $3.7$ & $0.2$ & $-2.44 \pm 0.07$ & $0.14 \pm 0.04$ & $0.23 \pm 0.04$ & $3.60 \pm 0.00$ \\
        HD144284 (FOCES) & $6280$ & $3.76$ & $0.11$ & $0.03$ & $0.06$ & $1.19$ & $0.16$ & $0.10 \pm 0.07$ & $0.14 \pm 0.00$ & $0.13 \pm 0.02$ & - \\
        HD162003 (FOCES) & $6608$ & $3.99$ & $-0.08$ & $0.1$ & $0.08$ & $1.69$ & $0.17$ & $-0.04 \pm 0.05$ & $0.15 \pm 0.15$ & $0.18 \pm 0.03$ & $1.79 \pm 0.00$ \\
        HD164259 (FOCES) & $6820$ & $4.03$ & $0.01$ & $0.08$ & $0.11$ & $1.24$ & $0.1$ & $-0.16 \pm 0.00$ & $0.20 \pm 0.00$ & $0.12 \pm 0.00$ & - \\
        HD220009 (NARVAL) & $4389$ & $1.83$ & $-0.72$ & $0.16$ & $0.4$ & $-0.21$ & $0.01$ & $-0.74 \pm 0.08$ & $0.25 \pm 0.07$ & $0.41 \pm 0.00$ & - \\
        HD22879 (NARVAL) & $5939$ & $4.27$ & $-0.93$ & $0.27$ & $0.48$ & $1.79$ & $0.15$ & $-0.88 \pm 0.09$ & $0.28 \pm 0.08$ & $0.44 \pm 0.03$ & $1.54 \pm 0.00$ \\
        HD284248 (FOCES) & $6293$ & $4.41$ & $-1.61$ & $0.29$ & $0.28$ & $3.22$ & $0.28$ & $-1.57 \pm 0.05$ & $0.30 \pm 0.09$ & $0.31 \pm 0.04$ & $2.99 \pm 0.00$ \\
        HD84937 (UVES) & $6397$ & $4.12$ & $-2.12$ & $0.36$ & $0.25$ & $3.2$ & $0.24$ & $-2.18 \pm 0.08$ & $0.35 \pm 0.07$ & $0.36 \pm 0.04$ & $3.36 \pm 0.00$ \\
        OMI\_AQL (FOCES) & $6162$ & $4.12$ & $0.15$ & $0.09$ & $0.06$ & $1.42$ & $0.6$ & $0.12 \pm 0.07$ & $0.06 \pm 0.11$ & $0.06 \pm 0.02$ & $1.49 \pm 0.00$ \\
        Procyon (NARVAL) & $6695$ & $3.9$ & $-0.06$ & $0.05$ & $0.16$ & $1.22$ & $0.23$ & $-0.06 \pm 0.08$ & $0.06 \pm 0.08$ & $0.21 \pm 0.06$ & - \\
        Procyon (UVES) & $6498$ & $3.64$ & $-0.14$ & $0.07$ & $0.16$ & $1.37$ & $0.06$ & $-0.16 \pm 0.07$ & $0.01 \pm 0.13$ & $0.26 \pm 0.04$ & - \\
        Procyon (UVES) & $6545$ & $3.73$ & $-0.13$ & $0.07$ & $0.21$ & $1.27$ & $0.12$ & $-0.14 \pm 0.06$ & $0.02 \pm 0.15$ & $0.23 \pm 0.05$ & - \\
        Sun (NARVAL) & $5746$ & $4.33$ & $-0.03$ & $0.11$ & $0.05$ & $0.32$ & $0.66$ & $-0.00 \pm 0.08$ & $0.12 \pm 0.04$ & $0.04 \pm 0.04$ & - \\
        Sun (UVES) & $5620$ & $4.17$ & $-0.21$ & $0.13$ & $0.15$ & $0.71$ & $0$ & $-0.16 \pm 0.04$ & $0.13 \pm 0.05$ & $0.13 \pm 0.02$ & - \\
        Sun (NARVAL) & $5768$ & $4.43$ & $-0.15$ & $0.1$ & $0.13$ & $0.79$ & $0$ & $-0.13 \pm 0.06$ & $0.09 \pm 0.05$ & $0.08 \pm 0.02$ & - \\
        $\alpha$ CenA (UVES) & $5590$ & $4.01$ & $-0.02$ & $0.13$ & $0.13$ & $0.14$ & $0$ & $0.03 \pm 0.06$ & $0.10 \pm 0.12$ & $0.17 \pm 0.05$ & - \\
        $\alpha$ Cet (NARVAL) & $3867$ & $0.71$ & $-0.52$ & $0.23$ & $0.05$ & $-1.19$ & $0.11$ & $-0.23 \pm 0.00$ & - & - & - \\
        $\alpha$ Cet (UVES) & $3791$ & $0.99$ & $-0.5$ & $0.13$ & $0.35$ & $-1.05$ & $0.05$ & - & - & - & - \\
        $\alpha$ Tau (NARVAL) & $3915$ & $0.99$ & $-0.4$ & $0.1$ & $0.25$ & $-1.2$ & $0$ & $-0.42 \pm 0.12$ & $0.10 \pm 0.02$ & - & - \\
        $\alpha$ Tau (UVES) & $3915$ & $0.99$ & $-0.44$ & $-0.01$ & $0.13$ & $-1.16$ & $0$ & $-0.58 \pm 0.19$ & $-0.17 \pm 0.02$ & - & - \\
        $\beta$ Hyi (UVES) & $5750$ & $3.65$ & $-0.26$ & $0.1$ & $0.19$ & $1.7$ & $0.03$ & $-0.23 \pm 0.08$ & $0.11 \pm 0.10$ & $0.19 \pm 0.01$ & $1.66 \pm 0.00$ \\
        $\beta$ Vir (NARVAL) & $6100$ & $3.96$ & $0.05$ & $0.07$ & $0.11$ & $1.02$ & $0.03$ & $0.03 \pm 0.08$ & $0.01 \pm 0.14$ & $0.10 \pm 0.03$ & $0.90 \pm 0.00$ \\
        $\delta$ Eri (UVES) & $4981$ & $3.63$ & $-0.12$ & $0.09$ & $0.08$ & $0.16$ & $0$ & $-0.17 \pm 0.08$ & $0.06 \pm 0.04$ & $0.13 \pm 0.02$ & - \\
        $\delta$ Eri (NARVAL) & $4994$ & $3.64$ & $-0.1$ & $0.11$ & $0.08$ & $0.15$ & $0$ & $-0.10 \pm 0.06$ & $0.08 \pm 0.07$ & $0.16 \pm 0.03$ & - \\
        $\delta$ Eri (UVES) & $4981$ & $3.7$ & $-0.12$ & $0.09$ & $0.16$ & $0.22$ & $0$ & $-0.16 \pm 0.07$ & $0.03 \pm 0.08$ & $0.11 \pm 0.02$ & - \\
        $\epsilon$ Eri (UVES) & $5017$ & $4.48$ & $-0.28$ & $0.13$ & $0.08$ & $0.08$ & $0$ & $-0.21 \pm 0.07$ & $0.14 \pm 0.06$ & $0.07 \pm 0.05$ & - \\
        $\epsilon$ Eri (UVES) & $5047$ & $4.61$ & $-0.25$ & $0.12$ & $0.12$ & $0.23$ & $0$ & $-0.24 \pm 0.08$ & $0.12 \pm 0.08$ & $-0.03 \pm 0.00$ & - \\
        $\epsilon$ Eri (UVES) & $4986$ & $4.48$ & $-0.28$ & $0.1$ & $0.12$ & $0.1$ & $0$ & $-0.25 \pm 0.07$ & $0.14 \pm 0.07$ & $-0.00 \pm 0.03$ & - \\
        $\epsilon$ Vir (NARVAL) & $5081$ & $2.75$ & $-0$ & $-0$ & $-0.02$ & $-0.06$ & $0.04$ & $-0.03 \pm 0.06$ & $0.01 \pm 0.06$ & $-0.02 \pm 0.02$ & - \\
        $\eta$ Boo (NARVAL) & $6133$ & $3.82$ & $0.22$ & $0.03$ & $0.08$ & $0.74$ & $0.07$ & $0.15 \pm 0.05$ & $0.00 \pm 0.12$ & $0.09 \pm 0.02$ & - \\
        $\mu$ Ara (UVES) & $5654$ & $4.11$ & $0.09$ & $0.11$ & $0.08$ & $0.38$ & $0$ & $0.08 \pm 0.06$ & $0.09 \pm 0.06$ & $0.10 \pm 0.02$ & - \\
        $\mu$ Cas (NARVAL) & $5328$ & $4.39$ & $-0.9$ & $0.35$ & $0.47$ & $1.16$ & $0.1$ & $-0.89 \pm 0.08$ & $0.34 \pm 0.05$ & $0.40 \pm 0.02$ & - \\
        $\mu$ Leo (NARVAL) & $4522$ & $2.45$ & $0.04$ & $0.03$ & $0.03$ & $-0.65$ & $0$ & $-0.03 \pm 0.07$ & $0.19 \pm 0.06$ & $0.14 \pm 0.06$ & - \\
        $\tau$ Cet (NARVAL) & $5404$ & $4.53$ & $-0.57$ & $0.27$ & $0.35$ & $0.92$ & $0.07$ & $-0.57 \pm 0.07$ & $0.23 \pm 0.04$ & $0.31 \pm 0.03$ & - \\
        \bottomrule
        \end{tabular}
        \begin{tablenotes}
            \item [$*$] Missing entries in the TSFitPy abundances are a result of too weak absorption line features in the considered wavelength range to perform a proper fit. 
        \end{tablenotes}
    \end{threeparttable}
\end{table*}

\begin{table*}[ht!]
    \centering
    \scriptsize
    \caption{Overview of raw cINN MAP predictions for benchmark spectra.}
    \label{tab:benchmark_predictions_raw}
    \begin{tabular}{llcccccccccccccccccccc}
    \toprule
    \multirow{2}{*}{Star (Spectrum)}&$T_\mathrm{eff}$ & \multirow{2}{*}{$\log\left(\frac{g}{\mathrm{cms}^{-2}}\right)$} & \multirow{2}{*}{$[\mathrm{Fe}/\mathrm{H}]$} & $v_\mathrm{broad}$ & $v_\mathrm{mic}$ & \multicolumn{8}{c}{$[\mathrm{x}/\mathrm{Fe}]$} & \multirow{2}{*}{$f_\mathrm{reject}$}\\
     & (K) &  &  & $\mathrm{kms}^{-1}$ & $\mathrm{kms}^{-1}$ & Ca & C & Li & Mg & Mn & N & Si & Ti & \\
    \midrule
    15PEG (FOCES) & $6415$ & $3.49$ & $-0.64$ & $11.14$ & $1.66$ & $0.08$ & $-0.51$ & $1.02$ & $0.21$ & $-0.26$ & $0.61$ & $-0.12$ & $-0.04$ & $0.13$ \\
    18Sco (NARVAL) & $5694$ & $4.22$ & $-0.14$ & $5.61$ & $1.2$ & $0.09$ & $-0.33$ & $0.46$ & $0.03$ & $-0.26$ & $0.55$ & $-0.12$ & $0.07$ & $0$ \\
    31AQL (FOCES) & $5488$ & $4.11$ & $0.2$ & $6.93$ & $1.27$ & $0.04$ & $0.06$ & $-0.12$ & $0.06$ & $-0.25$ & $0.64$ & $-0.09$ & $0.18$ & $0$ \\
    61CygA (NARVAL) & $4523$ & $4.65$ & $-0.39$ & $5.02$ & $0.54$ & $0.18$ & $0.14$ & $0.2$ & $0.23$ & $-0.19$ & $0.35$ & $-0$ & $0.4$ & $0.5$ \\
    61CygB (NARVAL) & $4288$ & $4.55$ & $-0.6$ & $5.4$ & $0.61$ & $0.42$ & $0.25$ & $0.02$ & $0.1$ & $-0.2$ & $0.6$ & $-0.04$ & $0.43$ & $0.72$ \\
    Arcturus (NARVAL) & $4415$ & $1.77$ & $-0.65$ & $6.94$ & $1.74$ & $0.17$ & $-0.1$ & $0.17$ & $0.39$ & $-0.16$ & $0.61$ & $-0.05$ & $0.59$ & $0$ \\
    Arcturus (UVES) & $4446$ & $1.89$ & $-0.61$ & $6.12$ & $1.73$ & $0.08$ & $-0.15$ & $0.27$ & $0.33$ & $-0.24$ & $0.49$ & $-0.07$ & $0.57$ & $0.01$ \\
    Arcturus (UVES) & $4435$ & $1.83$ & $-0.62$ & $6.39$ & $1.75$ & $0.12$ & $-0.2$ & $0.06$ & $0.41$ & $-0.19$ & $0.75$ & $-0.03$ & $0.57$ & $0.01$ \\
    BD+26\_3578 (FOCES) & $6379$ & $4.04$ & $-2.29$ & $7.75$ & $1.45$ & $0.35$ & $0.02$ & $3.11$ & $0.25$ & $-0.34$ & $0.2$ & $0.32$ & $0.22$ & $0.22$ \\
    BD-04\_3208 (FOCES) & $6274$ & $4.06$ & $-2.38$ & $6.79$ & $1.03$ & $0.28$ & $0.33$ & $3.05$ & $0.3$ & $-0.31$ & $0.33$ & $-0.02$ & $0.25$ & $0.16$ \\
    G113-22 (FOCES) & $5642$ & $3.65$ & $-1.18$ & $5.75$ & $1.41$ & $0.25$ & $-0.47$ & $1.84$ & $0.41$ & $-0.38$ & $0.51$ & $0.07$ & $0.29$ & $0.1$ \\
    G165-39 (FOCES) & $6304$ & $3.9$ & $-2.06$ & $8.69$ & $1.29$ & $0.32$ & $-0.23$ & $2.64$ & $0.2$ & $-0.19$ & $0.66$ & $0.16$ & $0.16$ & $0.17$ \\
    G207-5 (FOCES) & $5835$ & $4.11$ & $-0.55$ & $7.96$ & $1.12$ & $0.03$ & $-0.51$ & $1.2$ & $0.12$ & $-0.41$ & $0.76$ & $-0.1$ & $0.06$ & $0.36$ \\
    G28-43 (FOCES) & $5099$ & $4.63$ & $-1.72$ & $9.1$ & $1.19$ & $0.32$ & $-0.01$ & $2$ & $0.32$ & $-0.56$ & $0.81$ & $0.24$ & $0.32$ & $0.35$ \\
    G41-41 (FOCES) & $6395$ & $3.92$ & $-2.8$ & $2.76$ & $1.46$ & $0.13$ & $0.29$ & $3.49$ & $0.23$ & $-0.36$ & $0.39$ & $0.35$ & $0.05$ & $0.22$ \\
    G48-29 (FOCES) & $6395$ & $4.43$ & $-2.89$ & $3.78$ & $1.48$ & $0.24$ & $0.17$ & $3.37$ & $0.29$ & $-0.5$ & $0.46$ & $0.35$ & $0.28$ & $0.3$ \\
    G63-46 (FOCES) & $5741$ & $4.05$ & $-0.83$ & $5.65$ & $1.01$ & $0.22$ & $-0.45$ & $0.81$ & $0.36$ & $-0.25$ & $0.86$ & $0.02$ & $0.2$ & $0.23$ \\
    G64-12 (FOCES) & $6290$ & $4.67$ & $-3.77$ & $1.22$ & $1.12$ & $0.25$ & $0.32$ & $4.62$ & $0.64$ & $-0.47$ & $0.22$ & $0.43$ & $0.48$ & $0.3$ \\
    G78-1 (FOCES) & $5903$ & $3.94$ & $-0.97$ & $7.82$ & $1.32$ & $0.22$ & $-0.2$ & $1.66$ & $0.39$ & $-0.22$ & $0.93$ & $0.01$ & $0.25$ & $0.29$ \\
    G9-16 (FOCES) & $6764$ & $4.1$ & $-1.49$ & $12.96$ & $1.3$ & $0.33$ & $-0.27$ & $2.19$ & $0.32$ & $-0.55$ & $0.18$ & $0.22$ & $0.39$ & $0.13$ \\
    Gmb1830 (NARVAL) & $5044$ & $4.58$ & $-1.43$ & $2.67$ & $1.05$ & $0.35$ & $0.08$ & $1.53$ & $0.37$ & $-0.45$ & $0.75$ & $0.12$ & $0.36$ & $0.18$ \\
    HD101177 (FOCES) & $5829$ & $4.1$ & $-0.36$ & $7.15$ & $1.25$ & $0.05$ & $-0.46$ & $1.17$ & $0.11$ & $-0.23$ & $0.55$ & $-0.15$ & $0.08$ & $0.02$ \\
    HD102158 (FOCES) & $5650$ & $3.89$ & $-0.54$ & $5.84$ & $1.04$ & $0.26$ & $-0.28$ & $0.96$ & $0.36$ & $-0.37$ & $0.88$ & $-0.05$ & $0.21$ & $0.28$ \\
    HD103095 (FOCES) & $5080$ & $4.64$ & $-1.43$ & $3.18$ & $1.25$ & $0.33$ & $-0.1$ & $1.47$ & $0.36$ & $-0.41$ & $0.59$ & $0.15$ & $0.36$ & $0.23$ \\
    HD104304 (FOCES) & $5436$ & $4.4$ & $0.04$ & $7.12$ & $1.23$ & $0.07$ & $0.11$ & $0.03$ & $0.06$ & $-0.12$ & $0.38$ & $-0.07$ & $0.16$ & $0$ \\
    HD105631 (FOCES) & $5327$ & $4.49$ & $-0.07$ & $6.17$ & $1.22$ & $0.09$ & $-0.1$ & $0.36$ & $0.04$ & $-0.15$ & $0.57$ & $-0.14$ & $0.14$ & $0$ \\
    HD106210 (FOCES) & $5583$ & $4.17$ & $-0.3$ & $6.88$ & $1.16$ & $0.13$ & $-0.08$ & $0.81$ & $0.24$ & $-0.16$ & $0.23$ & $-0.03$ & $0.21$ & $0$ \\
    HD106516 (FOCES) & $6259$ & $4.01$ & $-0.81$ & $9.99$ & $1.53$ & $0.18$ & $-0.55$ & $1.48$ & $0.49$ & $-0.27$ & $0.67$ & $0.06$ & $0.14$ & $0.24$ \\
    HD10697 (FOCES) & $5517$ & $3.76$ & $-0.07$ & $7.86$ & $1.24$ & $0.04$ & $-0.11$ & $0.83$ & $-0$ & $-0.08$ & $0.44$ & $-0.17$ & $0.11$ & $0$ \\
    HD10700 (FOCES) & $5283$ & $4.4$ & $-0.67$ & $5.64$ & $1.07$ & $0.23$ & $-0.15$ & $0.86$ & $0.36$ & $-0.13$ & $0.67$ & $0$ & $0.3$ & $0.05$ \\
    HD107328 (UVES) & $4561$ & $2.15$ & $-0.53$ & $7.03$ & $1.84$ & $0.08$ & $-0.31$ & $0.24$ & $0.25$ & $-0.13$ & $0.44$ & $-0.09$ & $0.53$ & $0$ \\
    HD107328 (NARVAL) & $4541$ & $2.06$ & $-0.55$ & $6.79$ & $1.81$ & $0.1$ & $-0.08$ & $0.12$ & $0.34$ & $-0.03$ & $0.51$ & $-0.07$ & $0.52$ & $0$ \\
    HD110897 (FOCES) & $5791$ & $4.02$ & $-0.68$ & $6.97$ & $1.36$ & $0.08$ & $-0.28$ & $0.94$ & $0.21$ & $-0.31$ & $0.65$ & $-0.1$ & $0.04$ & $0.04$ \\
    HD111395 (FOCES) & $5490$ & $4.29$ & $-0.11$ & $7.42$ & $1.14$ & $0.1$ & $-0.34$ & $0.35$ & $0.01$ & $-0.13$ & $0.47$ & $-0.1$ & $0.13$ & $0$ \\
    HD113139 (FOCES) & $6817$ & $3.91$ & $-0.02$ & $91.06$ & $1.29$ & $0.06$ & $-0.29$ & $0.9$ & $-0.06$ & $-0.2$ & $0.49$ & $-0.01$ & $0.01$ & $0.08$ \\
    HD114174 (FOCES) & $5628$ & $4.15$ & $-0.14$ & $6.83$ & $1.22$ & $0.08$ & $-0.16$ & $0.11$ & $0.1$ & $-0.22$ & $0.36$ & $-0.13$ & $0.08$ & $0$ \\
    HD122563 (NARVAL) & $4871$ & $1.51$ & $-2.33$ & $4.56$ & $1.53$ & $0.1$ & $0.17$ & $2.13$ & $0.1$ & $-0.58$ & $0.04$ & $0.2$ & $0.25$ & $0.22$ \\
    HD140283 (NARVAL) & $5810$ & $3.58$ & $-2.42$ & $4.01$ & $1.12$ & $0.08$ & $0.2$ & $3.43$ & $0.18$ & $-0.21$ & $0.31$ & $0.37$ & $0.2$ & $0.2$ \\
    HD140283 (UVES) & $5923$ & $3.85$ & $-2.36$ & $2.7$ & $1.45$ & $0.06$ & $0.05$ & $3.28$ & $0.19$ & $-0.46$ & $0.27$ & $0.48$ & $0.22$ & $0.21$ \\
    HD144284 (FOCES) & $6179$ & $3.52$ & $0.06$ & $31.15$ & $1.67$ & $-0.01$ & $-0.45$ & $0.72$ & $-0.02$ & $-0.26$ & $0.93$ & $-0.14$ & $-0.06$ & $0.16$ \\
    HD162003 (FOCES) & $6478$ & $3.69$ & $-0.15$ & $15.32$ & $1.77$ & $0.04$ & $-0.68$ & $1.07$ & $-0$ & $-0.17$ & $0.75$ & $-0.23$ & $-0.07$ & $0.17$ \\
    HD164259 (FOCES) & $6671$ & $3.69$ & $-0.06$ & $71.31$ & $1.49$ & $0.01$ & $-0.51$ & $0.51$ & $0.03$ & $-0.33$ & $0.52$ & $-0.02$ & $0.01$ & $0.1$ \\
    HD220009 (NARVAL) & $4453$ & $1.96$ & $-0.72$ & $6.21$ & $1.48$ & $0.18$ & $-0.18$ & $0.23$ & $0.37$ & $-0.28$ & $0.44$ & $-0.05$ & $0.53$ & $0.01$ \\
    HD22879 (NARVAL) & $5867$ & $4.09$ & $-0.98$ & $4.43$ & $1.35$ & $0.23$ & $0.01$ & $1.48$ & $0.41$ & $-0.27$ & $0.81$ & $0.05$ & $0.27$ & $0.15$ \\
    HD284248 (FOCES) & $6190$ & $4.17$ & $-1.66$ & $5.59$ & $1.32$ & $0.25$ & $-0.51$ & $2.75$ & $0.21$ & $-0.51$ & $0.63$ & $0.04$ & $0.02$ & $0.28$ \\
    HD84937 (UVES) & $6285$ & $3.85$ & $-2.18$ & $3.47$ & $1.02$ & $0.31$ & $-0.36$ & $2.68$ & $0.17$ & $-0.61$ & $0.38$ & $0.23$ & $0.24$ & $0.24$ \\
    OMI\_AQL (FOCES) & $6071$ & $3.9$ & $0.09$ & $7.23$ & $1.31$ & $0.05$ & $-0.47$ & $1.01$ & $-0.01$ & $-0.15$ & $0.93$ & $-0.19$ & $0.02$ & $0.6$ \\
    Procyon (NARVAL) & $6557$ & $3.58$ & $-0.12$ & $8.93$ & $1.92$ & $-0.01$ & $-0.59$ & $0.55$ & $0.07$ & $-0.19$ & $0.49$ & $-0.17$ & $-0.07$ & $0.23$ \\
    Procyon (UVES) & $6378$ & $3.35$ & $-0.21$ & $8.95$ & $1.79$ & $0.01$ & $-0.6$ & $0.8$ & $0.07$ & $-0.26$ & $0.66$ & $-0.16$ & $-0.08$ & $0.06$ \\
    Procyon (UVES) & $6421$ & $3.44$ & $-0.2$ & $8.51$ & $1.82$ & $0.02$ & $-0.58$ & $0.67$ & $0.13$ & $-0.26$ & $0.57$ & $-0.15$ & $-0.08$ & $0.12$ \\
    Sun (NARVAL) & $5692$ & $4.2$ & $-0.07$ & $4.24$ & $0.92$ & $0.09$ & $-0.16$ & $0.11$ & $-0.01$ & $-0.09$ & $0.99$ & $-0.16$ & $0.07$ & $0.66$ \\
    Sun (NARVAL) & $5711$ & $4.29$ & $-0.19$ & $5.54$ & $1.23$ & $0.08$ & $-0.18$ & $0.57$ & $0.07$ & $-0.28$ & $0.64$ & $-0.11$ & $0.1$ & $0$ \\
    Sun (UVES) & $5576$ & $4.06$ & $-0.25$ & $6.11$ & $1.24$ & $0.11$ & $-0.15$ & $0.56$ & $0.09$ & $-0.22$ & $0.5$ & $-0.11$ & $0.09$ & $0$ \\
    $\alpha$ CenA (UVES) & $5549$ & $3.91$ & $-0.05$ & $6.74$ & $1.32$ & $0.11$ & $-0.01$ & $-0$ & $0.07$ & $-0.17$ & $0.4$ & $-0.04$ & $0.07$ & $0$ \\
    $\alpha$ Cet (UVES) & $3908$ & $1.24$ & $-0.48$ & $6.5$ & $1.65$ & $0.17$ & $-0.01$ & $-0.32$ & $0.33$ & $-0.18$ & $0.58$ & $-0.08$ & $0.48$ & $0.05$ \\
    $\alpha$ Cet (NARVAL) & $3977$ & $0.94$ & $-0.5$ & $6.67$ & $1.31$ & $0.28$ & $-0.14$ & $-0.5$ & $0.02$ & $-0.08$ & $0.56$ & $-0.09$ & $0.6$ & $0.11$ \\
    $\alpha$ Tau (NARVAL) & $4021$ & $1.21$ & $-0.39$ & $6.96$ & $1.59$ & $0.14$ & $-0.09$ & $-0.53$ & $0.22$ & $-0.13$ & $0.37$ & $-0.16$ & $0.43$ & $0$ \\
    $\alpha$ Tau (UVES) & $4021$ & $1.21$ & $-0.42$ & $5.15$ & $1.53$ & $0.03$ & $-0.34$ & $-0.5$ & $0.11$ & $-0.1$ & $0.32$ & $-0.2$ & $0.52$ & $0$ \\
    $\beta$ Hyi (UVES) & $5695$ & $3.51$ & $-0.3$ & $6.72$ & $1.34$ & $0.07$ & $-0.4$ & $1.48$ & $0.13$ & $-0.15$ & $0.54$ & $-0.16$ & $0.03$ & $0.03$ \\
    $\beta$ Vir (NARVAL) & $6014$ & $3.76$ & $-0.01$ & $7.38$ & $1.48$ & $0.03$ & $-0.36$ & $0.64$ & $0.04$ & $-0.21$ & $0.62$ & $-0.19$ & $-0.03$ & $0.03$ \\
    $\delta$ Eri (NARVAL) & $5005$ & $3.66$ & $-0.12$ & $5.85$ & $1.19$ & $0.12$ & $-0.06$ & $0.3$ & $0.04$ & $-0.11$ & $0.33$ & $-0.12$ & $0.26$ & $0$ \\
    $\delta$ Eri (UVES) & $4993$ & $3.64$ & $-0.13$ & $5.22$ & $1.26$ & $0.09$ & $0.09$ & $0.32$ & $0.03$ & $-0.2$ & $0.32$ & $-0.15$ & $0.25$ & $0$ \\
    $\delta$ Eri (UVES) & $4993$ & $3.71$ & $-0.14$ & $5.28$ & $1.27$ & $0.09$ & $0.04$ & $0.37$ & $0.11$ & $-0.12$ & $0.32$ & $-0.13$ & $0.25$ & $0$ \\
    $\epsilon$ Eri (UVES) & $4998$ & $4.5$ & $-0.3$ & $4.46$ & $1.14$ & $0.1$ & $-0.07$ & $0.25$ & $0.07$ & $-0.21$ & $0.49$ & $-0.08$ & $0.18$ & $0$ \\
    $\epsilon$ Eri (UVES) & $5053$ & $4.61$ & $-0.27$ & $3.91$ & $1.13$ & $0.12$ & $-0.23$ & $0.36$ & $0.07$ & $-0.23$ & $0.31$ & $-0.09$ & $0.22$ & $0$ \\
    $\epsilon$ Eri (UVES) & $5026$ & $4.48$ & $-0.3$ & $5.27$ & $1.14$ & $0.13$ & $-0.14$ & $0.22$ & $0.04$ & $-0.21$ & $0.46$ & $-0.07$ & $0.21$ & $0$ \\
    $\epsilon$ Vir (NARVAL) & $5084$ & $2.75$ & $-0.02$ & $7.58$ & $1.73$ & $-0$ & $-0.38$ & $0.05$ & $-0.07$ & $-0.15$ & $0.85$ & $-0.17$ & $0.21$ & $0.04$ \\
    $\eta$ Boo (NARVAL) & $6044$ & $3.61$ & $0.17$ & $14.88$ & $1.62$ & $-0.01$ & $-0.36$ & $0.34$ & $0$ & $-0.14$ & $0.65$ & $-0.18$ & $-0.01$ & $0.07$ \\
    $\mu$ Ara (UVES) & $5607$ & $3.99$ & $0.06$ & $6.4$ & $1.33$ & $0.09$ & $-0.17$ & $0.21$ & $0.02$ & $-0.19$ & $0.73$ & $-0.12$ & $0.1$ & $0$ \\
    $\mu$ Cas (NARVAL) & $5310$ & $4.34$ & $-0.93$ & $4.27$ & $0.97$ & $0.34$ & $-0.11$ & $1.15$ & $0.41$ & $-0.35$ & $0.74$ & $0.11$ & $0.33$ & $0.1$ \\
    $\mu$ Leo (NARVAL) & $4574$ & $2.55$ & $0.03$ & $7.34$ & $1.68$ & $0.05$ & $-0.03$ & $-0.28$ & $-0$ & $-0.19$ & $0.59$ & $-0.01$ & $0.41$ & $0$ \\
    $\tau$ Cet (NARVAL) & $5379$ & $4.46$ & $-0.6$ & $2.62$ & $0.99$ & $0.26$ & $-0.16$ & $0.87$ & $0.29$ & $-0.24$ & $0.6$ & $-0.04$ & $0.33$ & $0.07$ \\
    \bottomrule
    \end{tabular}
\end{table*}

\subsection{Bias correction}
\label{app:Bias_corr}
\begin{figure*}
    \centering
    \includegraphics[width=0.85\linewidth]{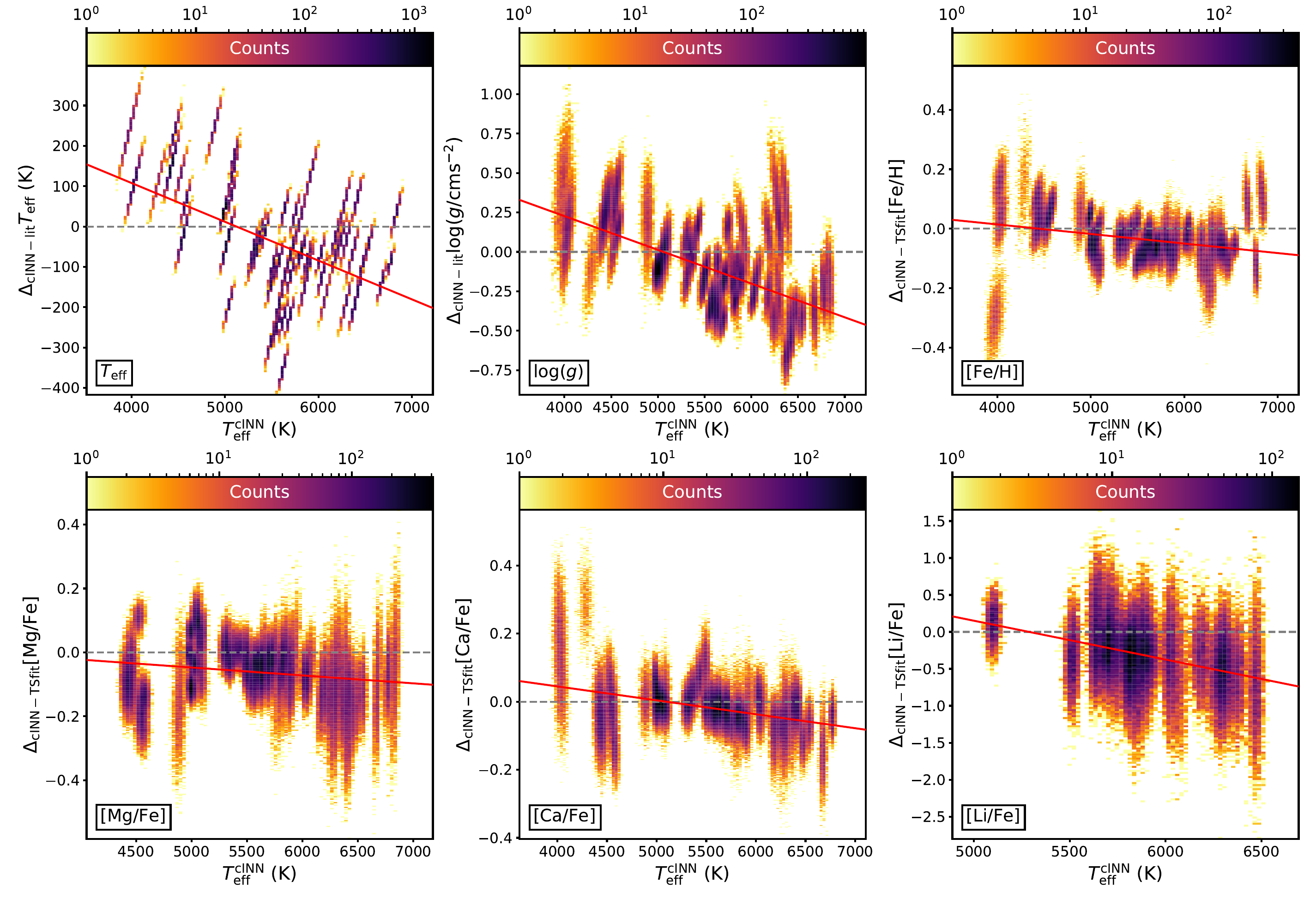}
    \caption{2D histogram of the residuals of the cINN posterior predictions with respect to the literature/TSFitPy results as a function of the estimates for \teff. The red lines indicate linear fits to the residuals used to correct for the apparent systematic bias in the cINN prediction.} 
    \label{fig:calibration_fit}
\end{figure*}

In our initial application to the set of observed spectra in Section~\ref{sec:results_benchmark}, we identified a small bias in the predicted stellar parameters and abundances in dependence of the \teff~estimates. This is illustrated in Fig.~\ref{fig:calibration_fit}, which shows a 2D histogram of the residuals of the cINN predictions with respect to the literature values for \teff~and \logg, and to the TSFitPy outcome for \feh, [Mg/Fe], [Ca/Fe] and [Li/Fe] as a function of the cINN estimates of \teff~for all benchmark spectra. For each star we plot all valid posterior samples in this comparison (i.e.~all samples remaining after the sample rejection procedure outlined in Sect.~\ref{sec:point_estimation}). To account for this systematic effect, we performed a linear correction, following
\begin{equation}
    \label{eq:lin_corr}
    \tilde{X}_\mathrm{cINN} = X_\mathrm{cINN} - (a_\mathrm{X, cal} T_\mathrm{eff, cINN} + b_\mathrm{X, cal}),
\end{equation}
where $a_\mathrm{X, cal}$ and $b_\mathrm{X, cal}$ refer to the slope and intercept of the linear fit (indicated in red in Fig.~\ref{fig:calibration_fit}) for target parameter X, and $T_\mathrm{eff, cINN}$ is the cINN estimate of the effective temperature.
We note that, following our assessment of the optimal performance range in our evaluation on the synthetic data in Sect.~\ref{sec:results_synthetic}, we excluded all stars from these fits that have a TSFitPy metallicity of $\mathrm{Fe/H}<-2.5$. Table~\ref{tab:calib_coeff} summarises the respective coefficients for all six target parameters considered in this analysis. We also note that we did not correct \vbroad~or \vmic, as reliable literature/ground truth values were not available for these parameters to compute the respective residuals.

\begin{table*}
    \centering
    \caption{Coefficients for linear correction of systematic offset of cINN predictions.}
    \label{tab:calib_coeff}
    \begin{tabular}{lcclcc}
    \toprule
    X & $a_\mathrm{X, cal}$ & $b_\mathrm{X, cal}$ & X & $a_\mathrm{X, cal}$ & $b_\mathrm{X, cal}$ \\
    \midrule
    \teff & -0.096 & $492.389$ & $[\mathrm{Mg/Fe}]$ & $-2.479 \times 10^{-5}$ & $0.077$ \\
    \logg & $-2.139 \times 10^{-4}$ & $1.082$ & $[\mathrm{Ca/Fe}]$ & $-4.095 \times 10^{-5}$ & $0.209$ \\
    \feh & $-3.209 \times 10^{-5}$ & $0.143$ & $[\mathrm{Li/Fe}]$ & $-5.254 \times 10^{-4}$ & $2.779$ \\
    \bottomrule
    \end{tabular}
\end{table*}

\subsection{Re-simulation}
\begin{figure*}
    \centering
    \includegraphics[width=\linewidth]{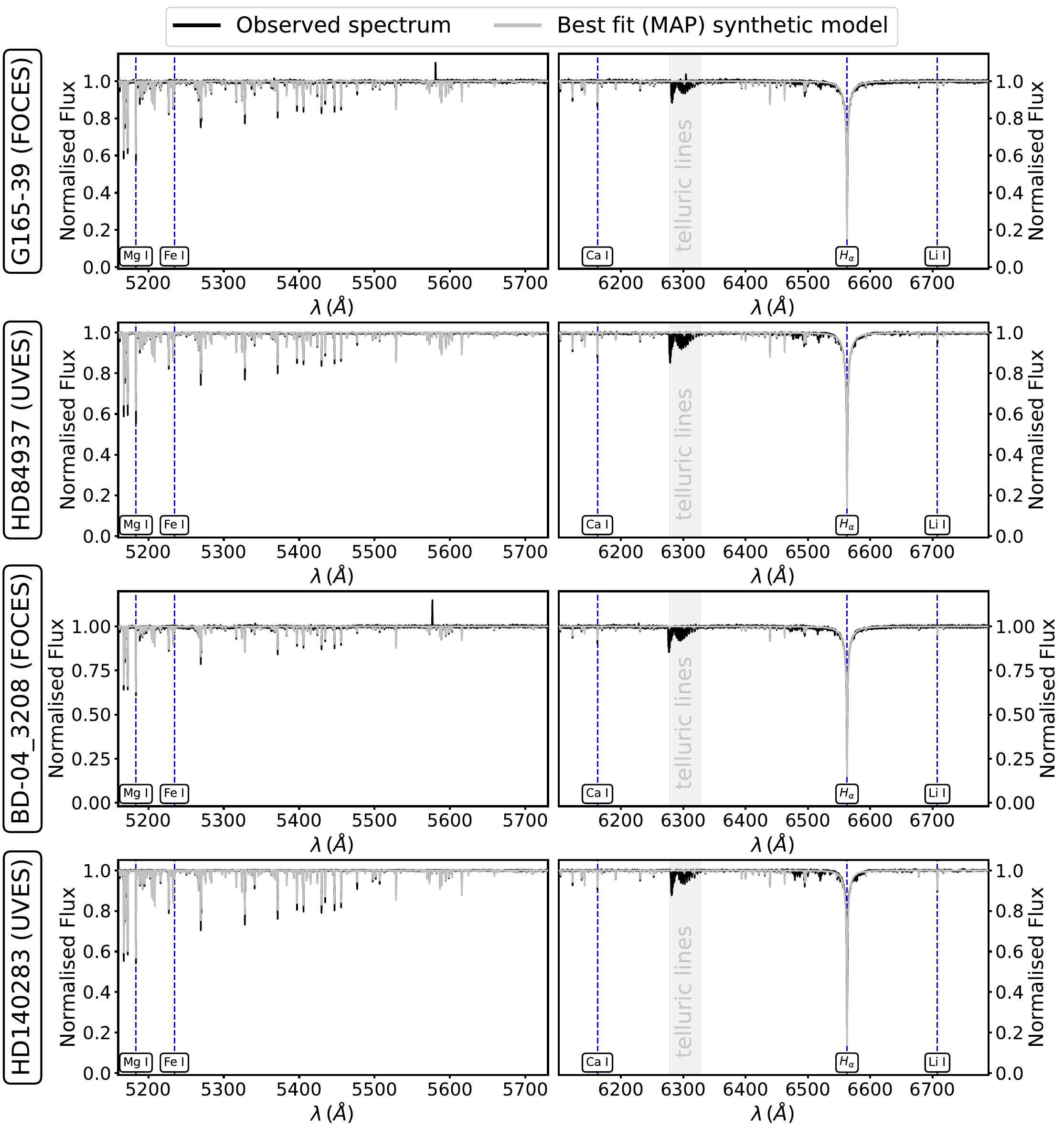}
    \caption{Comparison of the 4MOST-ified input spectra to model spectra corresponding to the cINN MAP estimates for four low-metallicity stars. The highlighted Mg, Fe, Li and Ca absorption lines correspond to the zoom-ins shown in Fig.~\ref{fig:resim_Mg_Ca_Li}.}
    \label{fig:resim_greenredwindow}
\end{figure*}

Figure~\ref{fig:resim_greenredwindow} provides additional results from our re-simulation experiment in Sect.~\ref{sec:results_benchmark}, showing the entire re-simulated spectra of the example stars shown in Fig.~\ref{fig:resim_Mg_Ca_Li} in comparison to the input observations. This diagram also indicates the position of the four example absorption lines that are the focus in Fig.\ref{fig:resim_Mg_Ca_Li}, the $H_\alpha$ absorption line and the telluric absorption features that have not been corrected for in the input observations. Not surprisingly, some of the more apparent discrepancies visible in these plots correspond to these telluric features at around 6300\,\AA~(see Fig.~\ref{fig:resim_greenredwindow}), as they are not being simulated in Turbospectrum.

\end{document}